\documentclass[a4paper,fleqn]{cas-sc}

\usepackage[authoryear]{natbib}
\usepackage{amssymb}
\usepackage{amsmath}
\usepackage{graphicx}
\usepackage{amsthm}
\usepackage{rotating}
\usepackage{adjustbox}
\usepackage{caption}
\usepackage{subcaption}
\usepackage{bm}
\usepackage{siunitx}
\usepackage{mathtools}
\usepackage{braket}
\usepackage{booktabs}
\usepackage{xcolor}

\newcommand{\ra}[1]{\renewcommand{\arraystretch}{#1}}
\newcommand{\ve}[1]{\mathbf{#1}}
\newcommand{\mat}[1]{\mathbf{#1}}
\newcommand{\vecnu}{\ve{\nu}}
\newcommand{\veceta}{\ve{\eta}}
\newcommand{\matJ}{\mat{J}}

\DeclareMathOperator{\diag}{diag}
\DeclareMathOperator{\sign}{sgn}
\DeclareMathOperator{\sig}{sig}

\newtheorem{assumption}{Assumption}
\newtheorem{theorem}{Theorem}[section]
\newtheorem{remark}{Remark}
\newtheorem{lemma}[theorem]{Lemma}
\newtheorem{definition}[theorem]{Definition}

\def\tsc#1{\csdef{#1}{\textsc{\lowercase{#1}}\xspace}}
\tsc{WGM}
\tsc{QE}
\tsc{EP}
\tsc{PMS}
\tsc{BEC}
\tsc{DE}
\begin{document}
\let\WriteBookmarks\relax
\def\floatpagepagefraction{1}
\def\textpagefraction{.001}

\shorttitle{}

\shortauthors{Hanzhi Yang et~al.}

\title [mode = title]{Fixed-time control with prescribed performance for path following of underwater gliders}                      
% \tnotemark[1,2]

% \tnotetext[1]{This document is the results of the research
%   project funded by the National Science Foundation.}

% \tnotetext[2]{The second title footnote which is a longer text matter
%   to fill through the whole text width and overflow into
%   another line in the footnotes area of the first page.}

\author[1]{Hanzhi Yang}[]

% \cormark[1]


\ead{yang1118@purdue.edu}

% \ead[url]{www.cvr.cc, cvr@sayahna.org}

\credit{Conceptualization, Data curation, Formal analysis, Investigation, Methodology, Software, Visualization, Writing - original draft}

\affiliation[1]{organization={Purdue University},
    addressline={School of Mechanical Engineering, Purdue University}, 
    city={West Lafayette},
    % citysep={}, % Uncomment if no comma needed between city and postcode
    postcode={47907 IN}, 
    country={USA}}

\author[1]{Nina Mahmoudian}[]
\cormark[1]
\ead{ninam@purdue.edu}
\credit{Conceptualization, Funding acquisition, Project administration, Resources, Supervision, Validation, Writing - review \& editing}

\cortext[cor1]{Corresponding author. School of Mechanical Engineering, Purdue University, West Lafayette, 47907 IN, USA}

% \fntext[fn1]{This is the first author footnote. but is common to third
%  author as well.}
%\fntext[fn2]{Another author footnote, this is a very long footnote and
%  it should be a really long footnote. But this footnote is not yet
%  sufficiently long enough to make two lines of footnote text.}

% \nonumnote{This note has no numbers. In this work we demonstrate $a_b$
%  the formation Y\_1 of a new type of polariton on the interface
%  between a cuprous oxide slab and a polystyrene micro-sphere placed
%  on the slab.
%  }

\begin{abstract}
Underwater gliders are increasingly deployed in challenging missions—such as hurricane-season observations and long-endurance environmental monitoring—where strong currents and turbulence pose significant risks to navigation safety. To address these practical challenges, this paper presents a fixed-time prescribed performance control scheme for the 3D path following of underwater gliders subject to model uncertainties and environmental disturbances. The primary contribution is the integration of a finite-time performance function within a fixed-time control framework. This synthesis ensures that the tracking errors are constrained within prescribed performance bounds and converge to a compact set within a fixed time, independent of initial conditions. A second key contribution is the development of a fixed-time sliding mode disturbance observer that provides accurate finite-time estimation of lumped disturbances, enhancing the system's robustness. Integrated with an iLOS guidance law, the proposed controller enables precise and safe waypoint following. Numerical simulations demonstrate that the proposed method outperforms conventional sliding mode and prescribed performance controllers in tracking accuracy, convergence speed, and control effort smoothness, validating its efficacy for robust underwater navigation. 
\end{abstract}

% \includegraphics{figs/grabs.pdf}

\begin{highlights}
\item A novel fixed-time prescribed performance controller is proposed, guaranteeing transient and steady-state tracking error converge within a fixed time independent of initial conditions.
\item The proposed controller is combined with a fixed-time sliding mode disturbance observer to achieve exact finite-time estimation of model uncertainties and environmental disturbances.
\item The integrated control scheme with line-of-sight guidance enables robust and accurate 3D path following for underwater gliders, outperforming conventional methods in simulations.
\end{highlights}

% Keywords

\maketitle

\section{Introduction}\label{Introduction}
%The Earth's oceans span 71\% of its surface yet have limited accessibility for scientific research. Unmanned underwater vehicles (UUVs) are widely used to conduct ocean explorations. 
 {Underwater gliders (UGs) are widely used in ocean observation missions such as hurricane forecasting, ecosystem and fishery monitoring, and water-quality supervision, because of their low energy consumption and resulting long endurance. However, operating in dynamic ocean environments poses significant challenges as currents, waves, and turbulence can threaten mission safety and degrade navigation performance. For example, more than 44 UG missions conducted by the National Oceanic and Atmospheric Administration (NOAA) in 2024 took place during hurricane seasons (\citet{UGexample}). Owing to their inherently low speed, UGs are highly susceptible to ocean currents (\citet{current_impact}), making it critical to develop robust control systems to ensure safe and reliable navigation.}

Motion control of UGs has been a trending topic for years. Most researchers divide the problem into vertical motion control and heading control. \citet{leonard2002model} linearized the UG model and applied the linear quadratic regulator (LQR) method to track the target attitudes. \citet{PID2} divided the UG system model into several single-input single-output subsystems and used a PID controller for flight angles and turning rates. \citet{Cao2016Adaptive} proposed an adaptive backstepping controller for sawtooth- and helix-shaped gliding trajectories and achieved smoother control performance than LQR. \citet{Song2017ADRC} applied active disturbance rejection control (ADRC) to improve the precision of pitch angle control in gliding motions. Compared to attitude controls, path tracking control has a characteristic of stronger comprehensiveness and needs to control multiple states at the same time (\citet{Wang2022Review}). Due to the high nonlinearity and underactuated features of UG dynamics, path tracking control has high requirements for the controller design, especially regarding the strong coupling states of the model. \citet{abraham2015model} used the model predictive control (MPC) method and time-pause technique to track gliding paths. \citet{zhang2014three} designed a robust $H_\infty$ controller to track helical trajectories in 3D space. Yet, these conventional control methods mainly focus on the steady state of UG motions, and the transient of the vehicle between steady glides has been neglected. The performance of such transient may include the convergence rate of the tracking errors and overshoot of the output, which are important for practical applications of UG. 

To control the transient performance of nonlinear systems, the prescribed performance control (PPC) method was first proposed by \citet{PPC1st} to constrain the performance of the tracking errors with a prescribed performance function (PPF) as the upper and lower bounds. The method uses an error transformation strategy to turn the constrained error dynamics into an equivalent unconstrained one, and thus facilitates the following controller design procedure. Applications of PPC on marine vehicle motion controls, such as \citet{PPCexample6,PPCexample7,PPCexample4}, have shown the promising enhancement of the method compared to conventional schemes. Furthermore, the finite-time performance function (FTPF) method, building upon the concept of PPC, enables presetting the convergence time for the tracking errors (\citet{FTPFearly}), and thus is effective in providing fast convergence time and strong robustness for marine vehicles. Recent studies like \citet{UGPPC,yang2025UGPPC,Luo2025UGPPC} applied FTPF on UG controls and accomplished better tracking performance in both vertical plane motion and heading angle controls compared with traditional robust control methods like sliding mode control (SMC). These studies concentrated on either plane trajectory tracking or helical spiral maneuvers instead of path tracking, but they highlight the potential of using FTPF to design a robust navigation controller for UGs to perform effective navigation in complex, dynamic water environments. 

Additionally, the fixed-time control (FxTC) method guarantees the tracking error convergence within a known bound, irrespective of the system's initial conditions. \citet{gao2020fixed} proposed a fixed-time sliding mode controller for AUV formation tracking and maintaining in a horizontal plane. \citet{su2021event} combined FxTC with an event-triggered integral sliding mode controller for trajectory tracking of AUVs in 3D space. \citet{wang2023predictor} designed a fixed-time heading controller and a fixed-time line-of-sight (LOS) guidance law for path following of unmanned surface vehicles (USVs) while using a predictor to estimate the environmental disturbances. \citet{Luo2025UGPPC} developed a fixed-time backstepping control scheme for UG attitude tracking in 3D space, providing firm control performance of sawtooth- and helix-shaped trajectory tracking. However, there are currently few works applying the FxTC method to UG's path following control, which enables more flexible maneuvers of the vehicle and yet is a challenging research area. 

UGs also face problems of model uncertainties and environmental disturbances. The dynamics of UGs are highly sensitive to the hydrodynamic coefficients (\citet{Wang2022Review}), the precise estimation of which is challenging, and thus, the controller's robustness can be hard to guarantee without considering the model uncertainties. In addition, in a dynamic ocean environment, disturbances like currents, waves, and turbulence can cause drifting in the trajectory and attitude of the vehicles, which may lead to mission failure. To address these issues, advanced techniques have been developed to mitigate the lumped disturbances. \citet{zhang2019fixed} developed a fixed-time extended state observer to approximate the lumped disturbances of USVs for trajectory tracking control. \citet{UGPPC} applied a radial basis function (RBF) neural network to estimate unknown smooth nonlinear disturbances for robust vertical motion control of UGs. \citet{yang2025UGPPC} designed a fixed-time sliding mode disturbance observer for UG heading control. 

Based on the above works, this paper focuses on the path following problem of UGs in a 3D space while maintaining prescribed performance, considering the model uncertainties and environmental disturbances. We combined the concepts of FxTC and FTPF to develop a novel control system that guarantees fast convergence time regardless of the initial conditions and limits the tracking error transient within a prescribed bound. Furthermore, we used a fixed-time disturbance observer to estimate the time-varying lumped disturbances, including the model uncertainties and environmental disturbances. Cooperating with an integral line-of-sight (iLOS) guidance law, we provide a control scheme for 3D path tracking of UGs in a complex, dynamic water environment. The main contributions of this paper are as follows: 
\begin{enumerate}
\item  {This work studied the prescribed performance control problem of UGs with a fixed-time stability by combining the concepts of FxTC and FTPF together. It also studies the UG path following problem in 3D space, in contrast to the works by \citet{PPCexample6,PPCexample4,UGPPC}. The proposed method provides accurate target tracking with low output oscillation and control effort chattering compared with conventional methods like SMC. In addition, we applied the fixed-time method to design a sliding mode disturbance observer that guarantees accurate convergence of estimating the unknown time-varying lumped disturbances in finite time. This further enhances the robustness of UG navigations in complex environments. }
\item By combining iLOS guidance law and the proposed fixed-time control law, we developed a comprehensive control system for UGs to track not only simply switching attitudes (like \citet{UGPPC,Luo2025UGPPC}) but also waypoint-based paths, providing a scheme that enables the gliders to perform more flexible and more complex maneuvers, which therefore may lead to more robust and reliable underwater navigations of UGs. 
\end{enumerate}

This paper is structured as follows. In Section \ref{Preliminaries}, notation and preliminary concepts are introduced. In Section \ref{Problem formulation}, the model of UGs and the control objective of this work are presented. In Section \ref{Control system design}, a fixed-time robust controller based on a novel FTPF is introduced. In addition, the fixed-time sliding mode disturbance observer is introduced, and iLOS guidance is added to the control system. In Section \ref{Results}, the proposed control method is compared with traditional SMC and the robust PPC method previously proposed in \citet{yang2025UGPPC} through numerical simulations in both periodic attitude switching and waypoint-based path following. Section \ref{Conclusion} summarizes the paper and discusses future work. 

%%%%%%%%%%%%%%%%%%%%%%%%%%%%%%%%%%%%%%%%%%%%%%%%%%%%%%%%%%
\section{Preliminaries}\label{Preliminaries}
\subsection{Notations}\label{notations}
The following notations are used in this paper: 
\begin{itemize}
    \item For a vector $\mathbf{v} \in \mathbb{R}^n$, $|\mathbf{v}|=[|v_1|, |v_2|, ..., |v_n|]^T$; 
    %\item For diagonal matrices $\ve{M_i}=\diag\{m_{i,1},...,m_{i,n}\}\in\mathbb{R}^{n\times n}$, $\min(\ve{M}_1,\ve{M}_2)=\diag\{\min(m_{1,1},m_{2,1}),...,\min(m_{1,n},m_{2,n})\}$.  
    \item The notations $\bullet^*$ and $\Delta\bullet$ are used to represent the actual value and the model uncertainty of a system parameter $\bullet$, s.t. its modeled value is $\bullet=\bullet^*-\Delta\bullet$. 
    \item $\diag\{\bullet\}$ is used as a vector-to-matrix operator is $\bullet$ is a vector, and as a matrix-to-vector operator if $\bullet$ is a diagonal matrix. 
    \item For the trigonometric functions in the matrices, this paper uses: $\mathrm{s}\bullet $ as $\sin{\bullet}$,  $ \mathrm{c} \bullet $  as $\cos{\bullet}$, $\mathrm{t}\bullet$ as $\tan{\bullet}$, and $\mathrm{sc} \bullet$ as $\sec{\bullet}$. 
    \item For variables $x,y\in\mathbb{R}$, the notation $\sig$ is a function defined as $\sig^x(y)=|y|^x\sign(y)$, and for vectors $\ve w,\mathbf{z}\in\mathbb{R}^n$, $\sig^\ve w(\mathbf{z}) = [sig^{w_1}(z_1),..., sig^{w_n}(z_n)]^T$. 
\end{itemize}
\begin{remark}
    From the definition of the $\sig$ function, it can be seen that $\frac{d}{dy}\sig^x(y)=x|y|^{x-1}$, $\sig^0(y)=\sign(y)$, $\sig^1(y)=y$, and $\sig^2(y)=y|y|$. 
\end{remark}

\subsection{Definitions and lemmas}
Consider a nonlinear system
\begin{equation}\label{eq: nlfunc}
    \dot{\ve x}(t)=f(\ve x(t)),\quad t>t_0, \quad \ve x(t_0)=\ve x_0
\end{equation}
in which $\ve x = [x_1, ..., x_n]^T\in\mathbb{R}^n$ is the state variable and $f(\ve x):\:\mathbb{R}^n\rightarrow\mathbb{R}^n$ is a nonlinear function. It is assumed that the origin is an equilibrium point of the system, and in this paper, it is assumed that $t_0=0$ with $x_0$ as the initial condition. 

\begin{definition}
    (\citet{Levant2005}). The origin of system \eqref{eq: nlfunc} is said to be globally finite-time stable if (1) it is Lyapunov stable, and (2) for any $R>0$ there exists $T>0$ such that any trajectory starting within the space $||\ve x||\leq R$ reaches to origin in the time $T$. 
\end{definition}
\begin{definition}
    (\citet{Polyakov2012}). The origin of system \eqref{eq: nlfunc} is said to be globally fixed-time stable if (1) it is globally finite-time stable and (2) there exists a fixed positive constant $T_{max}$ such that $T\leq T_{max}$ for any $\ve x_0\in\mathbb{R}^n$. 
\end{definition}
\begin{lemma}\label{lemma1}
    (\citet{Tian2017}). The convergence time $T$ for fixed-time stable systems is bounded even when the initial condition $\ve x_0$ tends to infinity. 
\end{lemma}
\begin{lemma}\label{lemma2}
    (\citet{gao2020fixed}). For a nonlinear system \eqref{eq: nlfunc}, if there exists a Lyapunov function that satisfies
    \begin{equation}
        \dot{V}(x)\leq -(\alpha V^p(x)+\beta V^q(x))^k
    \end{equation}
    with $\alpha,\beta, p, q, k\in\mathbb{R}>0$ and $pk<1,\:qk>1$, then the origin of the system is globally fixed-time stable, and its settling time $T$ is bounded by
    \begin{equation}
        T(\ve x_0) \leq T_{max} = \frac{1}{\alpha^k(1-pk)}+\frac{1}{\beta^k(qk-1)}, \:\forall \ve x_0\in\mathbb{R}^n
    \end{equation}
\end{lemma}
\begin{lemma}\label{lemma3}
    (\citet{gao2020fixed}). For a nonlinear system \eqref{eq: nlfunc}, if there exists a Lyapunov function that satisfies
    \begin{equation}
        \dot{V}(x)\leq -(\alpha V^p(x)+\beta V^q(x))^k+\vartheta
    \end{equation}
    with $\alpha,\beta, p, q, k\in\mathbb{R}>0$, $pk<1,\:qk>1$, and $\vartheta\in(0,+\infty)$ is finite, then the system is practical fixed-time stable and the residual set of its solution is
    \begin{equation}
        \mathcal{X}=\Set{\lim_{t\rightarrow T}\ve x|\|\ve x\|\leq\min\{\alpha^{-\frac{1}{p}}(\frac{\vartheta}{1-\theta})^{\frac{1}{p}}, \beta^{-\frac{1}{q}}(\frac{\vartheta}{1-\theta})^\frac{1}{q}\}}
    \end{equation}
    where $\theta\in(0,1)$, and the settling time $T$ is bounded by
    \begin{equation}
        T\leq T_{max}:=\frac{1}{\alpha\theta(1-p)}+\frac{1}{\beta\theta(q-1)}
    \end{equation}
    
\end{lemma}

%%%%%%%%%%%%%%%%%%%%%%%%%%%%%%%%%%%%%%%%%%%%%%%%%%%%%%%%%%
\section{Problem formulation}\label{Problem formulation}
 {This section provides the system model of UGs considering the lumped disturbances including modeling uncertainties and external disturbances. The control objective has two aspects: to design a fixed-time controller for the UG model to track the reference attitudes, and to combine a path following mechanism so that the reference positions can be reached in finite time and thus the preset course can be followed. }
\subsection{Model of underwater gliders}
This paper takes the SeaWing glider (\citet{zhang2013spiraling}) as the major object for modeling. Definitions of the parameters used in the equations are listed in Table. \ref{tab: notations}. The glider system model is here rewritten in the form of Fossen's model (\citet{fossen1994guidance}) following the steps in \citet{yang2024underice} with the following assumptions: 
\begin{assumption}\label{assumption1}
Because the sideslip velocities $v,w$ are usually infinitesimal during equilibrium flights, it is assumed that $v$, $w$ $\ll$ u, s.t. $V=u$, $\cos{\alpha}=1$, $\sin{ \alpha}=\frac{w}{u}$, $\cos{\beta}=1$, $\sin{\beta}=\frac{v}{u}$. 
\end{assumption}
\begin{assumption}\label{assumption2}
    For control input $\gamma$, the first three degrees of approximation by Taylor expansion are taken, s.t.  $\cos{\gamma}=1-\frac{\gamma^2}{2}+\frac{\gamma^4}{24}$, $\sin{\gamma}=\gamma-\frac{\gamma^3}{6}+\frac{\gamma^5}{120}$. This provides reliable estimates of the trigonometric terms for $|\gamma|\leq\frac{\pi}{2}$ with estimate errors $|\epsilon|<\frac{\pi}{150}$. 
\end{assumption}
\begin{assumption}\label{assumption3}
Because the model uncertainties and the environmental disturbances are usually caused by modeling errors and ocean currents, they are assumed to be bounded with known boundaries, s.t. $|\ve d|\leq \ve d_M$ and $|\dot{\ve d}|\leq\dot{\ve d}_M$, where $\ve d_M=[d_{M,1}, ..., d_{M,6}]^T$ and $\dot{\ve d}_M=[\dot{d}_{M,1}, ..., \dot{d}_{M,6}]^T$ are known positive constant vectors.  
\end{assumption}

\begin{figure}
    \centering
    \includegraphics[width=.5\linewidth]{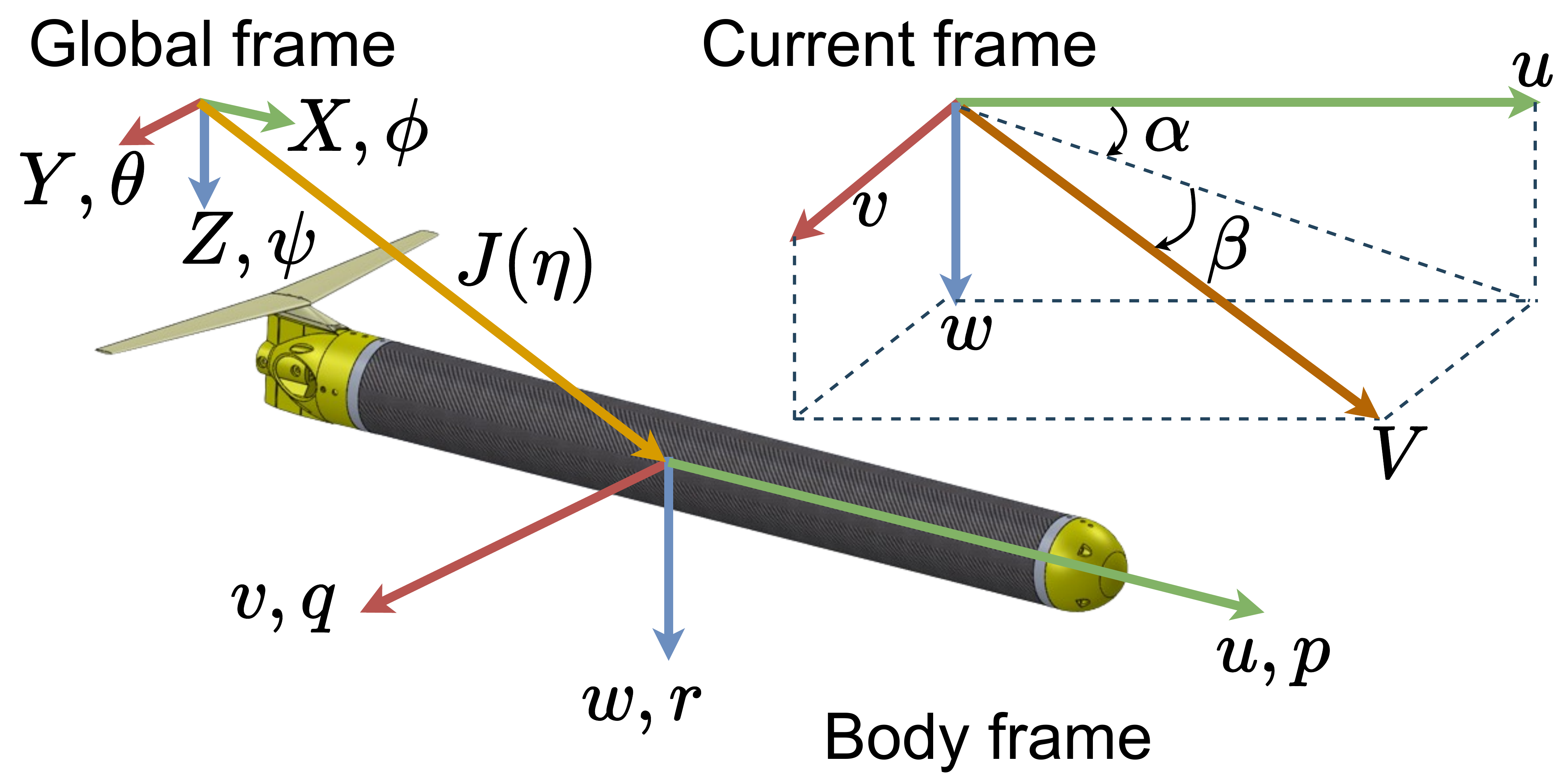}
    \caption{ {NED coordinate system for underwater gliders. }}
    \label{fig: Coord}
\end{figure}

The mathematical model of UGs is described in a body-fixed coordinate frame and a global coordinate frame as shown in Fig. \ref{fig: Coord}. The kinematic equation is given as 
\begin{equation} \label{eq: eom-kinematics2}
    \dot{\veceta} = \matJ(\veceta)\vecnu
\end{equation}
and the dynamic equation including environmental disturbances is expressed as
\begin{equation}
    \label{eq: eom-dynamics2}
   \mat M^*\dot{\ve\nu} = \mat C^*(\ve\nu)\ve\nu+\mat D^*(\ve\nu)\ve\nu+\mat B^*(\ve\eta,\ve\nu)\ve U+\mat E^*(\ve\eta)+\ve{\tau}_d
\end{equation}
where \(\veceta = [Z,\theta,\psi]^T\) is the vehicle's depth, pitch angle, and heading angle in the global frame, \(\vecnu = [u,v,w,p,q,r]^T \) is the vehicle's velocity in the body frame. The input \(\ve U=[m_b,r_{p_1},\gamma]^T\), where $m_b$ is buoyancy mass, $r_{p_1}$ is the position of the moving mass on the longitudinal axis in the body frame, and $\gamma$ is the angular position of the moving mass from the gravity and buoyancy vector. \(\ve{\tau}_d=[\tau_{d,1},...,\tau_{d,6}]^T\) is the external disturbance in the body frame, mainly including the influence of underwater environments such as waves and water currents. 

\begin{table}[t]
    \caption{Definitions of parameters in the underwater glider model}
    \centering
    \begin{adjustbox}{max width=\linewidth}
    \begin{tabular}{||c|c||} 
    \hline
    \textbf{Notation} & \textbf{Definition} \\ [0.5ex] 
    \hline\hline
         $m_1, m_2, m_3$ & Mass including added terms \\ 
         \hline
         $I_1, I_2, I_3$     & Moment of inertia including added terms\\ 
         \hline
        $X,Y,Z$ & Positions in global frame\\
        \hline
        $\phi,\theta,\psi$ & Roll/pitch/yaw angles \\
        \hline
        $u,v,w$ & Velocities in body frame\\
        \hline
        $p,q,r$ & Angular velocities \\ 
        \hline 
        $\alpha, \beta$ & Angles of attack and slip\\
        \hline 
        $K^*$ & Hydrodynamic coefficients \\ 
        \hline 
        $g$ & Gravity coefficient\\ 
        \hline 
        $m_b, m_p, m_h$ & Buoyancy/moving/hull mass \\
        \hline 
        $R_p$ & Offsets of moving mass\\
        \hline
        $r_b,r_p$ & Positions of buoyancy/moving mass\\
        \hline
        $\gamma$ & Rotation angle of moving mass\\
        \hline
        
    \end{tabular}
    \end{adjustbox}
    
    \label{tab: notations}

\end{table}

In the equations of motion \eqref{eq: eom-kinematics2}, $\matJ(\ve\eta)$ is the rotation matrix expressed as
\begin{equation}
    \matJ = \begin{bmatrix}
            \mat T & \mat\varnothing_{1\times3} \\ 
            \mat\varnothing_{2\times3} & \mat R
      \end{bmatrix}
\end{equation}
where 
\begin{equation}
    \mat T=\begin{bmatrix}
        -\mathrm s\theta & \mathrm s\phi \,\mathrm c\theta & \mathrm c\phi \,\mathrm c\theta
    \end{bmatrix}
\end{equation}
\begin{equation}
    \mat R =\begin{bmatrix}
        0 & \,\mathrm c\phi & -\mathrm s\phi \\
        0 & \,\mathrm s\phi \,\mathrm {sc}\theta & \,\mathrm c\phi \,\mathrm {sc}\theta
    \end{bmatrix}
\end{equation}
\(\mat M\) is the inertia matrix including added terms, 
\begin{equation}
    \label{eq: eom-M}
    \mat M^* = \diag\{m_1,m_2,m_3,I_1,I_2,I_3\}
\end{equation}
\(\mat C\) is the Coriolis matrix for which
\begin{equation}
    \label{eq: eom-Cv}
    \mat C^*\vecnu = \begin{bmatrix}
        m_2vr-m_3wq\\
        -m_1ur+m_3wp\\
        m_1uq-m_2vp\\
        m_2vw-m_3wv+I_2qr-I_3rq\\
        -m_1uw+m_3wu-I_1pr+I_3rp\\
        m_1uv-m_2vu+I_1pq-I_2qp
    \end{bmatrix}
\end{equation}
\(\mat D\) is the external hydrodynamic matrix for which
\begin{equation}
    \label{eq: eom-Dv}\mat D^*\vecnu = \begin{bmatrix}
        K_{L_0}uw-K_{D_0}u^2\\
        K_{\beta}uv-K_{D_0}uv\\
        -K_{L_0}u^2-K_{D_0}uw-K_Lwu\\
        K_ppu^2+K_{mr}uv-K_{m_0}uv\\
        K_{m_0}u^2+K_qqu^2+K_muw\\
        K_rru^2+K_{my}uv
    \end{bmatrix}
\end{equation}
\(\mat B\ve U+\mat E\) separates the internal actuation control inputs from the gravity and buoyancy matrix and is given by

\begin{equation}
    \label{eq: eom-B}
     \begin{adjustbox}{max width=\linewidth}
    $ \mat B^* = \begin{bmatrix}
        -g\cdot \mathrm s\theta & 0 & 0 \\
        g\cdot \mathrm c\theta \,\mathrm s\phi & 0 & 0 \\
        g\cdot \mathrm c\theta \,\mathrm c\phi & 0 & 0 \\
        0 & 0 & -m_pR_pg\mathrm c\theta[\mathrm c\phi(1-\gamma^2/6+\gamma^4/120)-\mathrm s\phi(-\gamma/2+\gamma^3/24)] \\ 
        -g\cdot \mathrm c\phi \,\mathrm c\theta r_b  &  -m_pg \mathrm c\phi \,\mathrm c\theta  & -m_pgR_p\mathrm s\theta(-\gamma/2+\gamma^3/24)\\ 
        g\cdot \mathrm c\theta \,\mathrm s\phi r_b &  m_pg \mathrm c\theta \,\mathrm s\phi & m_pgR_p\mathrm s\theta(1-\gamma^2+\gamma^4/120)
    \end{bmatrix}
    $
    \end{adjustbox}
\end{equation}
and
\begin{equation}
    \label{eq: eom-E}
    \mat E^* = \begin{bmatrix}
        0 \\
        0 \\
        0 \\
        -g\cdot \mathrm c\phi R_pm_p \\ 
        -g\cdot \mathrm s\theta R_pm_p\\ 
        0
    \end{bmatrix}
\end{equation}
Considering the model uncertainties, including unmodeled dynamics and uncertain hydrodynamic coefficients, the dynamics model \eqref{eq: eom-dynamics2} can be rewritten as 
\begin{equation}
    \mat M\dot{\vecnu}=\mat C\vecnu+\mat D\vecnu+\mat B\ve U+\mat E+\ve d
\end{equation}
where $\ve d$ represents the lumped disturbances, including model uncertainties and environmental disturbances, and is given by
\begin{equation}\label{eq: lumped disturbance}
    \ve d = -\Delta \mat M\dot{\vecnu}+\Delta \mat C\ve \nu+\Delta \mat D\ve\nu+\Delta \mat B\ve U+\Delta \mat E+\ve{\tau}_d
\end{equation}
\begin{remark}
    The uncertainty terms in the lumped disturbances are velocity-related functions, as shown in  \eqref{eq: lumped disturbance}, so it can further be assumed that their derivatives are also bounded by velocity-related functions. Plus, since the environmental disturbance term is limited, Assumption \ref{assumption3} is reasonable. 
\end{remark}

\subsection{Control objective}\label{objective}
The equations of motions in \eqref{eq: eom-kinematics2} and \eqref{eq: eom-dynamics2} can be rewritten as
\begin{equation}
    \ddot{\ve \eta}=\dot{\ve J}\ve\nu+\ve J\ve M^{-1}(\ve C\ve\nu+\ve D\ve \nu+\ve B\ve U+\ve E+\ve d)
\end{equation}
and thus in a compact format as
\begin{equation}\label{eq: eom}
    \ddot{\ve\eta}=\ve f+\ve g\ve U+\ve h\ve d
\end{equation}
where
\begin{align}
    \ve f &= \dot{\ve J}\ve\nu+\ve J\ve M^{-1}(\ve C\ve\nu+\ve D\ve\nu+\ve E)\\
    \ve g &= \ve J\ve M^{-1}\ve B\\
    \ve h &= \ve J\ve M^{-1}
\end{align}

 {This work focuses on the path tracking of UGs, which requires the control of depth, pitch angle, and heading angle. Therefore, the control objective is to design a fixed-time control law for the buoyancy mass and position of the moving mass that enables the vehicle, with its system model defined in \eqref{eq: eom}, to track time-varying target attitudes ($\eta_d$) including depth, pitch and heading in finite time with high robustness under the influence of model uncertainties and environmental disturbances.  A mathematical expression of the control objective is
\begin{equation}
    \lim_{t\rightarrow T_{max}} \eta-\eta_d=0
\end{equation}
By adding a guidance law, this controller can further enable a finite-time tracking of the target positions in 3D space ($[X_d,Y_d,Z_d]^T$) and thus follow the preset path.}

The control objective uses the following assumptions: 
\begin{assumption}\label{assumption4}
    The state variables of the UG system, including $\ve\eta$ and $\ve\nu$, are available as real-time feedback. 
\end{assumption}
\begin{assumption}\label{assumption5}
    The target attitude $\ve\eta_d$ is bounded, and its rate of change $\dot{\eta}_d, \ddot{\eta}_d$ are bounded and continuous. 
\end{assumption}

%%%%%%%%%%%%%%%%%%%%%%%%%%%%%%%%%%%%%%%%%%%%%%%%%%%%%%%%%%
\section{Control system design}\label{Control system design}
 {In this section, a fixed-time control that guarantees prescribed performance is designed. A fixed-time disturbance observer is added to the controller to enhance its robustness by estimating the lumped disturbances in finite time. The proposed control system is combined with the iLOS guidance method to enable waypoint-based path following. }
\subsection{Fixed-time disturbance observer}
 We used a fixed-time sliding mode disturbance observer to estimate the lumped disturbances that include environmental disturbances and model uncertainties of the underwater glider. {The observer introduces the estimated body-frame velocity $\ve\varpi$, whose derivative is given by
\begin{equation}\label{eq: SMDO2}
    \dot{\ve\varpi} = \mat M^{-1}(\mat C\vecnu+\mat D\vecnu+\mat E-\ve\rho+\mat B\ve U-\int_0^t\ve \varphi \:d\tau)
\end{equation}
in which two variables are used to update the estimated velocity based on the estimate error ($\Pi$), 
\begin{equation}\label{eq: SMDO3}
    \ve \rho=-\ve \iota_1(\sig^{\frac{1}{2}}(\ve \Pi)+\ve\varsigma\sig^{\frac{3}{2}}(\ve\Pi))
\end{equation}
\begin{equation}\label{eq: SMDO5}
    \ve\varphi=-\ve\iota_2(2\ve\varsigma\ve\Pi+\frac{3}{2}\ve\varsigma^2\sig^2(\ve\Pi)+\frac{1}{2}\sig^0(\ve\Pi))
\end{equation}
where the estimate error is given by
\begin{equation}\label{eq: SMDO1}
    \mat \Pi = \mat M\vecnu - \mat M\ve \varpi
\end{equation}
The estimate of lumped disturbances is found by
\begin{equation}\label{eq: SMDO4}
    \hat{\ve d} = -\int_0^t\ve\varphi d\tau
\end{equation}
}
For each degree of freedom in the body frame, the coefficients $\ve\iota_1$ and $\ve\iota_2$ can be chosen from the following set
    \begin{equation}\label{eq: SMDO6}
      \begin{split}
        \mathcal I_o = &\Set{(\iota_{1,i}, \iota_{2,i})\in\mathbb{R}^2| 0<\iota_{1,i}\leq2\sqrt{\dot{d}_{M,i}}, \iota_{2,i}>\frac{\iota_{1,i}^2}{4}+\frac{4\dot{d}_{M,i}^2}{\iota_{1,i}^2}} \\ 
        & \cup \Set{(\iota_{1,i}, \iota_{2,i})\in\mathbb{R}^2| \iota_{1,i}>2\sqrt{\dot{d}_{M,i}}, \iota_{2,i}>\dot{d}_{M,i}}
      \end{split}
    \end{equation}
where $i=1,2,3,4,5,6$.
\begin{theorem}
\label{theorem SMDO}
    With disturbance observer design as \eqref{eq: SMDO2}-\eqref{eq: SMDO4} and the coefficients chosen from the set in \eqref{eq: SMDO6}, the lumped disturbance $\ve d$ can be estimated by $\hat{\ve d}$ in a finite time $T_{obs}$. 
\end{theorem}
\begin{proof}
    See \citet{yang2025UGPPC}. 
\end{proof}

\subsection{Finite-time prescribed performance function}
According to the definition of PPF proposed in \citet{PPC1st}, a performance function $P(t)$ is a continuous positive function that is strictly monotone decreasing on $[0,+\infty)$. This method was introduced to constrain the tracking errors of a system within a specific range during the entire transient while balancing the overshoot and error convergence laws. Building upon the definition of PPF, \citet{FTPFearly} defines the FTPF as a smooth performance function $P(t)$ that satisfies (1) $\lim_{t\rightarrow T}P(t)=P_\infty$, (2) $P(t)=P_\infty \:\forall t\geq T$. This finite-time method enables the system to adjust and converge rapidly within a preset timeframe. This work utilizes a hyperbolic function to construct the FTPF as
\begin{equation}
    \label{eq: PPF}
    P(t) = \begin{cases}
        \mathrm{sech}(\mathrm{sech}(P_0)\cdot\frac{T}{T-t})+P_\infty, & 0\leq t<T \\ 
        P_\infty, &t\geq T
    \end{cases}
\end{equation}
where $\mathrm{sech}(\bullet)=2/[(]exp(\bullet)+exp(-\bullet)]$. $P_0$ and $P_\infty$ are positive constants satisfying $\mathrm{sech}(P_0)>P_\infty$. $T$ is a preset time constant, with which $P(t)$ reaches its minimum $P_\infty$ at $t=T$ and then maintains at the minimum $\forall t>T$. 
\begin{remark}
    The classical form of FTPF proposed in \citet{FTPFearly}  is given as
    \begin{equation}\label{eq: original FTPF}
        P(t) = \begin{cases}
            (P_0-\frac{t}{T})\exp{(1-\frac{T}{T-t})}+P_\infty ,& 0\leq t<T \\
            P_\infty, &t\geq T
        \end{cases}
    \end{equation}
Different from the exponential form of FTPF, this paper chooses the hyperbolic secant function to design the performance function. A comparison between the two types of FTPF is shown in Fig. \ref{fig: FTPF comparison}. The new FTPF provides a larger boundary for the tracking error at the beginning and a smoother transient process at the point where $t$ is closed to $T$. This design of FTPF gives slow-varying systems like UGs more space to adjust during the error convergence process ($ 0\leq t<T$) and less radical transient from converging to constant phase ($t\doteq T$). 

\begin{figure}
    \centering
    \includegraphics[width=.5\linewidth]{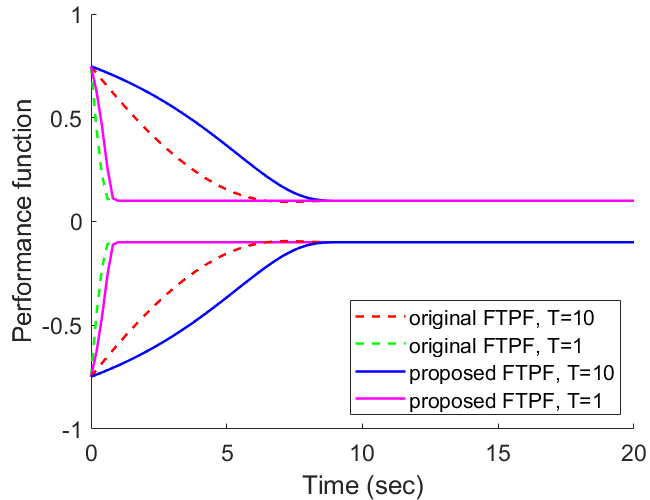}
    \caption{Comparison between the original FTPF and the proposed FTPF used in this work, with the same initial and final values for all functions and the different preset settling times.}
    \label{fig: FTPF comparison}
\end{figure}

\end{remark}

The attitude tracking error is defined as 
\begin{equation}\label{eq: error definition}
    \ve e = [e_1,e_2,e_3]^T = \ve\eta - \ve\eta_d
\end{equation}
and each error element is bounded by 
\begin{equation}\label{eq: constrained system}
    -\delta_{L,i}P_i(t)<e_i(t)<\delta_{R,i}P_i(t)
\end{equation}
where $\delta_{L,i}, \delta_{R,i}\in(0,1]$ for $i=1,2,3$ define the lower and upper overshoot boundaries of each tracking error. 

An error transformation function is designed to transform the constrained system in \eqref{eq: constrained system}  to an unconstrained one. Such a function is defined as 
\begin{equation}
    e_i(t) = P_i(t)S_i(\varepsilon_i)
\end{equation}
where $\varepsilon_i$ is the transformed error. The function $S_i(\varepsilon_i)$ is defined as
\begin{equation}
    S_i(\varepsilon_i)=\frac{\delta_{R,i}\exp(\varepsilon_i)-\delta_{L,i}\exp(-\varepsilon_i)}{\exp(\varepsilon_i)+\exp(-\varepsilon_i)}
\end{equation}
with which the transformed error can be found as
\begin{equation}\label{eq: transformed error}
    \varepsilon_i=\frac{1}{2}\ln(\frac{\delta_{L,i}+e_i/P_i}{\delta_{R,i}-e_i/P_i})
\end{equation}
Taking the second derivative of \eqref{eq: transformed error} yields the transformed error dynamics
\begin{equation}\label{eq: error dynamics 1}
   \dot{\varepsilon_i} = \lambda_i(\frac{\dot{e_i}P_i-e_i\dot{P_i}}{P_i})
\end{equation}
\begin{equation}\label{eq: error dynamics 2}
    \ddot{\varepsilon_i}=\dot{\lambda_i}(\frac{\dot{e_i}P_i-e_i\dot{P_i}}{P_i})+\lambda_i(\frac{\ddot{e_i}P_i^2-\dot{e_i}\dot{P_i}P_i-e_i\ddot{P_i}P_i+e_i\dot{P_i}^2}{P_i^2})
\end{equation}
where 
\begin{equation}
    \lambda_i=\frac{1}{2P_i}(\frac{1}{\delta_{L,i}+e_i/P_i}+\frac{1}{\delta_{R,i}-e_i/P_i})
\end{equation}
The transformed error dynamics \eqref{eq: error dynamics 1} and \eqref{eq: error dynamics 2} can be further written into a compact form by letting $\epsilon_{1,i}=\varepsilon_i$ and $\epsilon_{2,i}=\dot{\varepsilon}_i$, then plugging in \eqref{eq: eom} and \eqref{eq: error definition} yields
\begin{align}\label{eq: sliding surface error}
        \dot{\epsilon}_{1,i} &= \epsilon_{2,i} \\
        \dot{\epsilon}_{2,i} &= \mathbf{K} + \ve\Lambda(-\ddot{\ve\eta}_d+\ve f+\ve{gU}+\ve{hd})
\end{align}
where 
\begin{equation}
    \ve\Lambda=\diag\{\lambda_1,\lambda_2,\lambda_3\}
\end{equation}
and
\begin{align}
        \ve{K}&=[\kappa_1,\kappa_2,\kappa_3]^T\\
        \kappa_i&=\dot{\lambda}_i(\frac{\dot{e}_iP_i-e_i\dot{P}_i}{P_i})+\lambda_i(\frac{-\dot{e}_i\dot{P}_iP_i-e_i\ddot{P}_iP_i+e_i\dot{P}_i^2}{P_i^2})
\end{align}
\begin{remark}
    {To assist with controller tuning in practice, the following guidelines are proposed for selecting the parameters for the FTPF \eqref{eq: PPF}:
\begin{itemize}
    \item $P_0$: Set to $2-3 \times$ the expected initial error magnitude of each attitude to ensure early-stage flexibility;
    \item $P_\infty$: Chosen based on steady-state error tolerance, typically 5-10\% of  the maximum allowable deviation;
    \item $\delta_L, \delta_R$: Can be set to 1 for symmetric error bounds or reduced to tighten performance margins;
    \item $T$: Represents the maximum acceptable settling time, and should match mission-specific convergence needs (e.g., $30-60$ seconds for waypoint switching). 
\end{itemize}
 The parameter values are tuned to meet the desired performance of the glider. }
\end{remark}

\subsection{Fixed-time prescribed performance control law}
Define an integral sliding surface using the FTPF transformed errors \eqref{eq: sliding surface error}

\begin{equation}\label{eq: sliding surface}
\begin{split}
    \ve s=&\ve \epsilon_2+\ve k_1\int_0^t[\sig^{\ve \varrho_1}(\ve \epsilon_1)+\sig^1(\ve \epsilon_1)+\sig^{\ve \varrho_1'}(\ve \epsilon_1)]\:d\tau
    +\ve k_2\int_0^t[\sig^{\ve \varrho_2}(\ve \epsilon_2)+\sig^1(\ve \epsilon_2)+\sig^{\ve \varrho_2'}(\ve \epsilon_2)]\:d\tau
\end{split}
\end{equation}

where $\ve{k}_i=\diag\{k_{i,1},k_{i,2},k_{i,3}\}\in\mathbb{R}^{3\times3}>0$, $\ve\varrho_i=[\varrho_{i,1},\varrho_{i,2},\varrho_{i,3}]^T\in\mathbb{R}^3>0$ for $i=1,2$. The values of $\ve\varrho_i$'s are computed using the following equations:

\begin{equation}
\begin{cases}
    \varrho_1 & =\frac{\varrho}{2-\varrho}\\
    \varrho_1' & =\varrho \\ 
    \varrho_2 &= \frac{4-3\varrho}{2-\varrho} \\ 
    \varrho_2' &= \frac{4-3\varrho}{3-2\varrho}
\end{cases}
\end{equation}
where $\varrho\in(0,1)$. 
%The values of $\ve k_i$'s are chosen to ensure that $
%    s^2+k_{i,2}s+k_{i,1}$ and $ 
%    s^2+3k_{i,2}s+3k_{i,1}$ are Hurwitz. 
The fixed-time prescribed performance control law (FxTPPC) for 3D trajectory tracking is designed as {
\begin{equation}\label{eq: control law}
\ve U = \ve g^{-1}\bigl(-\mat\Lambda^{-1}\mathbf{ K}+\ddot{\mathbf{\eta}}_d-\ve f-\hat{\ve d} + \ve u_{\epsilon} + \ve u_{s}\bigr)
\end{equation}
in which the two control inputs based on the transformed errors yield
\begin{align}
    &\ve u_\epsilon = -\mat \Lambda^{-1}[\mat k_1(\sig^{\ve \varrho_1}(\ve \epsilon_1)+\sig^{1}(\ve \epsilon_1)+\sig^{\ve \varrho_1'}(\ve \epsilon_1))+\ve k_2(\sig^{\ve\varrho_2}(\ve\epsilon_2)+\sig^{1}(\ve\epsilon_2)+\sig^{\ve\varrho_2'}(\ve\epsilon_2))]\\
    &\ve u_s = -\mat \Lambda^{-1}[\mat k_1(\sig^{[1+1/\mu]}(\ve s)+\sig^{1}(\ve s)+\sig^{[1-1/\mu]}(\ve s))+\ve k_2(\sig^{[1+1/\mu]}(\ve s)+\sig^{1}(\ve s)+\sig^{[1-1/\mu]}(\ve s))]\}
\end{align}}
where $[1\pm1/\mu]=\begin{bmatrix}
    1\pm1/\mu_1 \\ 
    1\pm1/\mu_2 \\
    1\pm1/\mu_3
\end{bmatrix}$, $\mu_i>1$. 

\begin{theorem}\label{theorem controller}
    For a UG 3D trajectory tracking control problem, considering the lumped disturbances including model uncertainties and environmental disturbances, if the control system is designed as \eqref{eq: control law}, with properly chosen controller parameters, the tracking errors are globally fixed-time stable after a fixed time $T_{max}$, and therefore the prescribed performances can be satisfied in finite time. 
\end{theorem}
\begin{proof}
    For the control system \eqref{eq: control law}, define a Lyapunov function candidate: 
    \begin{equation}
        V = \sum_{i=1}^3V_i = \sum_{i=1}^3|s_i|
    \end{equation}
    which is the Euclidean 1-norm of the sliding surface \eqref{eq: sliding surface}, then its derivative yields
    \begin{align}\label{eq: proof}\allowdisplaybreaks
        \dot{V} =&\sum_{i=1}^3{\sign(s)\dot{s}}\notag\\
        =&\sum_{i=1}^3\{\sign(s)[\lambda_i(d_i-\hat{d_i})-k_{1i}(\sig^{1+1/\mu_i}(s_i)+\sig^1(s_i)+\sig^{1-1/\mu_i}(s_i))\notag\\
        &\quad\quad\quad\quad\quad\quad\quad\quad\quad\quad
        -k_{2i}(\sig^{1+1/\mu_i}(s_i)+\sig^1(s_i)+\sig^{1-1/\mu_i}(s_i))]\}\notag\\
        =&\sum_{i=1}^3\{\sign(s)\lambda_i\Tilde{d}_i-[k_{1i}(\sig^{1+1/\mu_i}(s_i)+\sig^1(s_i)+\sig^{1-1/\mu_i}(s_i))]\sign(s)\notag\\
        &\quad\quad\quad\quad\quad\quad\quad
        -[k_{2i}(\sig^{1+1/\mu_i}(s_i)+\sig^1(s_i)+\sig^{1-1/\mu_i}(s_i))]\sign(s)\}\notag\\
        \leq&\sum_{i=1}^3\{\lambda_i|\Tilde{d}_i|-k_{1i}(|s_i|^{1+1/\mu_i}+|s_i|^{1-1/\mu_i})\notag\\
        &\quad\quad\quad\quad\quad
        -k_{2i}(|s_i|^{1+1/\mu_i}+|s_i|^{1-1/\mu_i})\}\notag\\
        \leq&\sum_{i=1}^3\{\lambda_i|\Tilde{d}_i|-k_{1i}(|s_i|^{1+1/\mu_i}+|s_i|^{1-1/\mu_i})\notag\\
        &\quad\quad\quad\quad\quad
        -k_{2i}(|s_i|^{1+1/\mu_i}+|s_i|^{1-1/\mu_i})\}\notag\\
        \leq&\sum_{i=1}^3\lambda_i|\Tilde{{d}}_i| - \sum_{i=1}^3\{\min{(k_{1i},k_{2i})}(V_i^{1+1/\mu_i}+V_i^{1-1/\mu_i})\}%\notag\\
 %       \leq&\sum_{i=1}^3\lambda_i|\Tilde{{d}}_i| - \min{(k_{1,1},k_{1,2},...,k_{3,2})}\sum_{i=1}^3(V_i^{1+1/\mu_i}+V_i^{1-1/\mu_i})
    \end{align}
By Theorem \ref{theorem SMDO}, the disturbance estimate error $|\Tilde{\ve d}|$ can guarantee its convergence in a finite time $T_{obs}$, so for $t\in(T_{obs},+\infty)$, $|\Tilde{\ve d}|\rightarrow0$, and thus in the view of Lemma \ref{lemma2}, the fixed-time stability of the control system can be achieved after a finite time. For $t\in(0,T_{obs})$, since $\sum\lambda|\Tilde{d}|\in(0,+\infty)$, from Lemma \ref{lemma3}, it can be concluded that the system is practical fixed-time stable with the solution bounded within a residual set. From Eq. \eqref{eq: proof}, it can be seen that for each control objective, 
\begin{equation}
    V_i\leq \lambda_i|\Tilde{d}_i|-min(k_{1i},k_{2i})(V_i^{1+1/\mu_i}+V_i^{1-1/\mu_i})
\end{equation}
for $i=1,2,3$, which means that for individual sub-control systems (depth, pitch and heading), fixed-time stability can be achieved. According to Lemma \ref{lemma2}, the settling time for each control objective is bounded by 
\begin{equation}
    T_i(\eta_{i_0})\leq T_{i_{max}}=\frac{2\mu_i}{\min(k_{1_i},k_{2_i})}
\end{equation}
Therefore, the total convergence time of the sliding surface \eqref{eq: sliding surface} is bounded by
\begin{equation}
    T_m\leq T_{obs}+\sum_{i=1}^3 T_{i_{max}}
\end{equation}
The definition of the sliding surface also guarantees the transformed error \eqref{eq: sliding surface error} to be uniformly bounded, which proves that the prescribed performance \eqref{eq: constrained system} can be satisfied, and therefore the tracking errors are each bounded by their own performance functions \eqref{eq: PPF}. This completes the proof of Theorem \ref{theorem controller}. 
\end{proof}
\begin{remark}
    {For each control sub-objective (depth, pitch and heading), the tuning of the corresponding controller coefficients $k_{1,i}$ and $k_{2,i}$ may start with a small value of $k_{1,i}$, ($k_{2,i}$ can be automatically computed to satisfy the inequality $k_{1,i}<\frac{3}{4}k_{2,i}^2$,) and then gradually increase the value of $k_{1,i}$ to enhance the rate of convergence and to ensure the tracking error can be bounded by the performance functions for all control objectives.}
\end{remark}

\subsection{Path following guidance law}
The iLoS guidance law is utilized to calculate the target heading angles based on the position of the underwater glider in the horizontal plane; this enables the vehicle to either maintain its course in a straight path or actively adjust its heading angles to follow the waypoints. The guidance law is defined as in \citet{iLoS}: 
\begin{align}\label{eq: guidance law}
    &\psi_d=-\tan^{-1}(\frac{y_e+k_I\sigma_{int}}{\Lambda})\\
    &\dot{\sigma}_{int}=\frac{\Lambda y_e}{\Lambda^2+(y_e+k_I\sigma_{int})^2}
\end{align}
where $y_e$ is the cross-track error, $\Lambda\in\mathbb{R}>0$ is the Line-of-Sight distance, and $k_I\in\mathbb{R}>0$ is the integral gain, a parameter to be tuned. 

\begin{figure*}
    \centering
    \includegraphics[width=1\textwidth]{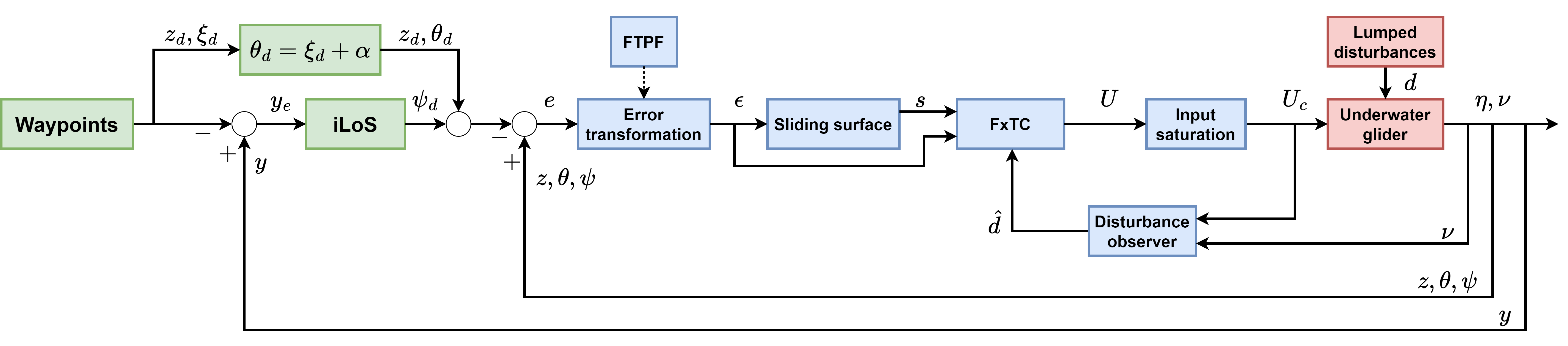}
    \caption{Block diagram of the proposed control system utilizing a fixed-time prescribed performance controller, a fixed-time sliding mode disturbance observer, and an integral LoS guidance law.}
    \label{fig: block diagram}
\end{figure*}

The position tracking in the vertical plane is achieved by following a series of target depth $z_d$ and desired gliding angle $\xi_d$, which gives the required pitch angle by $\theta_d=\xi_d+\alpha$. The overall control system for UGs path tracking in 3D space, combining the dynamic control law \eqref{eq: control law} and kinematic guidance law \eqref{eq: guidance law}, is shown as a block diagram in Fig. \ref{fig: block diagram}.

%%%%%%%%%%%%%%%%%%%%%%%%%%%%%%%%%%%%%%%%%%%%%%%%%%%%%%%%%%
\section{Numerical simulation results}\label{Results}
In this section, the control system proposed in this paper is tested and compared with the traditional SMC method and the PPC method proposed in \citet{yang2025UGPPC}.  The simulated underwater glider is a SeaWing glider (\citet{zhang2013spiraling}); its hydrodynamic parameters are listed in Table. \ref{tab: parameters} and are used as the actual system parameter values in the simulation.  A 20\% uncertainty is added to the model, and the time-varying environmental disturbances are simulated using sinusoid functions given as
\begin{equation}
    %d=0.1(-M^*\dot{\nu}+C^*\nu+D^*\nu+B^*\nu+E^*)+
    \ve\tau_d=\begin{bmatrix}
        0.02\sin(\frac{1}{100}\pi t)\\
        0.01\sin(\frac{1}{100}\pi t)\\
        -0.02\sin(\frac{2}{300}\pi t) \\ 
        0.01\sin(\frac{1}{100}\pi t)\\
        0.02\sin(\frac{1}{100}\pi t\\
        -0.01\sin(\frac{1}{100}\pi t)
    \end{bmatrix}
\end{equation}
The initial states of the glider are set to be $\eta(0)=\nu(0)=0$. The disturbance observer coefficients are set to be $\ve\iota_1=\diag\{0.001,0.001,0.01,0.01,0.01,0.01\}$, $\ve\iota_2=\diag\{0.01,0.01,0.1,0.1,0.1,0.1\}$, $\ve\varsigma=\diag\{18,18,18,180,180,180\}$. The proposed FxTPPC uses the following controller parameter values: $\ve \varrho=[0.8,0.8,0.8]^T$, $\ve\mu=[2,2,2]^T$, $\ve k_1=\diag\{0.001,0.01,0.001\}$, $\ve k_2=\diag\{0.01,0.2,0.08\}$. 

To verify the efficacy of the proposed control method, two baseline controllers, SMC and PPC, are used for comparison. The baseline controllers are designed as follows, 

\begin{itemize}
    \item \textbf{SMC}: The sliding surface is designed as 
    \begin{equation}\label{eq: SMC_ss}
        \ve p = \dot{\ve e}+\ve c_0\ve e
    \end{equation}
    where $\ve c_0\in\mathbb{R}^{3\times3}>0$ is a controller parameter to be designed. Then the SMC law is designed as 
    \begin{equation}
        \ve{U}_{smc} = \ve{g}^{-1}[-\ve{c}_0\dot{\ve e}+\ddot{\ve\eta}_d-\ve{f}-\ve{c}_1\ve{p}-\ve{c}_2\tanh{(\ve{p})}]
    \end{equation}
    where $\ve{c}_1, \ve{c}_2\in\mathbb{R}^{3\times3}>0$. In this work, the parameter values are tuned as $\ve{c}_0=\diag\{0.1,0.1,0.1\}$, $\ve{c}_1=\diag\{1,1,1\}$, $\ve{c}_2=\diag\{5,5,5\}$. Note that the hyperbolic tangent function is used to replace the conventional signum function for reducing the control effort chattering issue. 

    \item \textbf{PPC}: Here we expand the prescribed performance heading controller in \citet{yang2025UGPPC} to a 3D attitude tracking controller, the FTPF for each control objective is 
    \begin{equation}\label{eq: PPC FTPF}
        \mathcal{P}(t) = \begin{cases}
            \frac{\exp(\mathcal{P}_0\cdot T/(T-t))}{1+\exp^2(\mathcal{P}_0\cdot T/(T-t))}+\mathcal{P}_\infty, & 0\leq t<T \\ 
            \mathcal{P}_\infty, &t\geq T
        \end{cases}
    \end{equation}
    The transformed errors are then used to construct the sliding surface 
    \begin{equation}
        \ve{q} = \dot{\ve\varepsilon} + \ve{l}_0\varepsilon
    \end{equation}
    with controller parameter$\ve l_0\in\mathbb{R}^{3\times3}>0$. Then the PPC law yields
    \begin{equation}
        \ve U_{ppc} = \ve{g}^{-1}[-\ve{\Lambda}^{-1}\ve{\mathcal{\ve K}}+\ddot{\eta}_d-\ve{f}-\ve{l}_1\ve{q}-\ve{l}_2\tanh{(\ve{q})}-\hat{\ve d}]
    \end{equation}
    in which $\ve{l}_1, \ve{l}_2\in\mathbb{R}^{3\times3}>0$. The values of the parameters are selected as $\ve l_0=\diag\{0.1,0.1,0.1\}$, $\ve l_2=\diag\{1,1,3\}$, and $\ve l_1$ are automatically computed to ensure $\frac{d}{dt}{(\diag\{\ve\Lambda^{-1}\})}-\ve l_1<0$. 
\end{itemize}

\begin{table}[]
    \caption{System parameters of SeaWing glider}
    \centering
    \begin{adjustbox}{max width=\linewidth}
    \begin{tabular}{||c|c|c||} 
    \hline
    \textbf{Parameter} & \textbf{Notation} & \textbf{Value} \\ [0.5ex] 
    \hline\hline
         Moving mass  & $m_p$ & $11\:\si{kg}$ \\ 
         \hline
         Hull mass & $m_h$ & $54.28\:\si{kg}$ \\ 
         \hline 
         Offset of moving block & $R_p$ & $0.014\:\si{m}$ \\
         \hline 
         Buoyancy mass & $m_b$ & $\in[-0.4,0.4]\:\si{kg}$ \\ 
         \hline
         Position of moving mass & $r_{p_1}$ & $\in[-0.06,0.06]\:{\si{m}}$ \\
         \hline 
         Angle of moving mass & $\gamma$ & $\in[-\pi/2,\pi/2]\:{\si{rad}}$ \\
         \hline 
         Body mass & ${\mat M}=\diag\{m_1,m_2,m_3\}$ & $\diag\{66.76, 114.86, 131.20\}\:{\si{kg}}$ \\ 
         \hline
         Body moment of inertia & ${\mat I}=\diag\{I_1,I_2,I_3\}$ & $\diag\{1.13, 23.15, 25.50\}\:{\si{kg\cdot m^2}}$ \\
         \hline 
         Coefficients of drag force & $K_D, K_{D0}$ & $386.29, 7.19$ \\
         \hline
         Coefficients of lift force & $K_L,K_{L0}$ & $440.99, -0.36$  \\
         \hline
         Coefficient of side force & $K_\beta$ &  $-115.65$  \\
         \hline
         Coefficient of roll moment & $K_{MR}, K_p$ &  $-58.27, -19.83$  \\
         \hline
         Coefficient of pitch moment & $K_M,K_{M0},K_q$ &  $-65.84, 0.28, -205.64$  \\ 
         \hline 
         Coefficient of yaw moment  &  $K_{MY}, K_r$  &  $34.10, -389.30$  \\
        \hline
        
    \end{tabular}
    \end{adjustbox}
    
    \label{tab: parameters}

\end{table}

\subsection{Case 1: Periodic target attitude switching}\label{case1}
In this test case, the UG is commanded to track a series of reference depth, gliding angle, and heading angle in an 800-second period. The reference angles are set to be
\begin{align}\label{eq: reference gliding angle}
    \xi_d&=\begin{cases}
        -\pi/4, &0\leq t<200 \\
        -\pi/3, &t<400 \\
        -\pi/4, &t<600 \\
        -\pi/3, &t\leq800
    \end{cases} \\
    \psi_d&=\begin{cases}
        \pi/6, &0\leq t<200 \\
        0, &t<400 \\
        -\pi/6, &t<600 \\
        \pi/10, &t\leq800
    \end{cases}
\end{align}
The target depth is set with a constant rate of change
\begin{equation}\label{eq: reference depth}
    Z_d=0.1t
\end{equation}
and the desired pitch angle is computed in real-time as $\theta_d=\xi_d+\alpha$. 

When the target attitude is switched, the instant tracking error is significant. Therefore, a performance switching mechanism is added the proposed FTPF \eqref{eq: PPF}:
\begin{equation}
    \begin{split}
        t^*&=mod(t,200)\\
        P(t^*)& = \begin{cases}
        \mathrm{sech}(\mathrm{sech}(P_0)\cdot\frac{T}{T-t^*})+P_\infty, & 0\leq t^*<T \\ 
        P_\infty, &t^*\geq T
    \end{cases}
    \end{split}
\end{equation}
In this test case, the FTPF parameters are selected for each control objective:
\begin{itemize}
    \item \textbf{$Z$} \text{--} $P_0=1$, $P_\infty=0.2$, $T=100$, $\delta_L=1$, $\delta_R=1$;
    \item \textbf{$\theta$} \text{--} $P_0=5\pi/18$, $P_\infty=\pi/18$, $T=80$, $\delta_L=1$, $\delta_R=1$;
    \item \textbf{$\psi$} \text{--} $P_0=5\pi/18$, $P_\infty=\pi/12$, $T=100$, $\delta_L=1$, $\delta_R=1$. 
\end{itemize}
The performance function of PPC is set to have the same initial value ($\mathcal{P}_0$), final value ($\mathcal{P}_\infty$), and finite time ($T$) as the setups shown above.

\begin{figure}
    \centering
    \includegraphics[width=.45\linewidth]{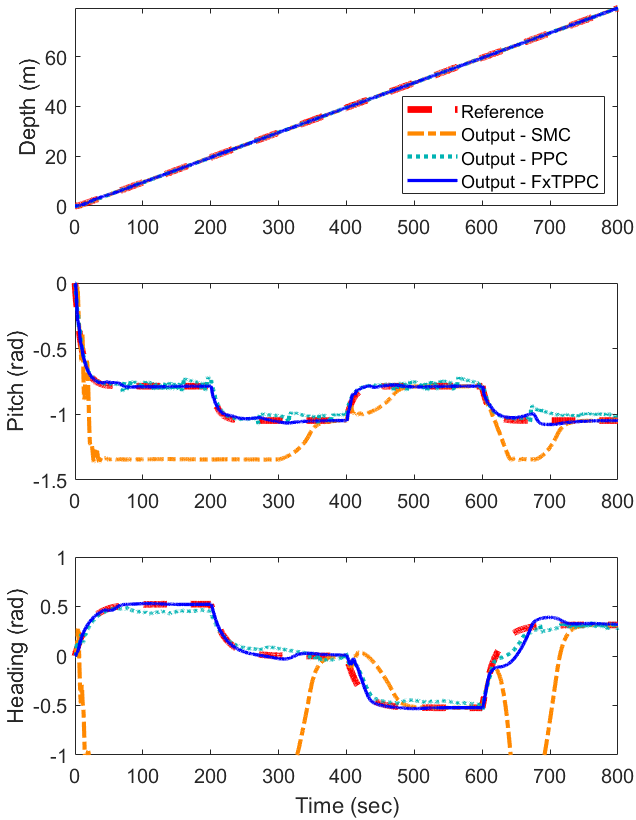}
    \caption{{Switching attitude tracking performance comparison among SMC, PPC, and the proposed FxTPPC. In the presence of lumped disturbances, SMC results in large drifting from the reference and PPC leads to high output chattering while FxTPPC is able to track the reference with high accuracy and low oscillation.}}
    \label{fig: switch tracking performance}
\end{figure}

\begin{figure*}
    \centering
    \includegraphics[width=.8\textwidth]{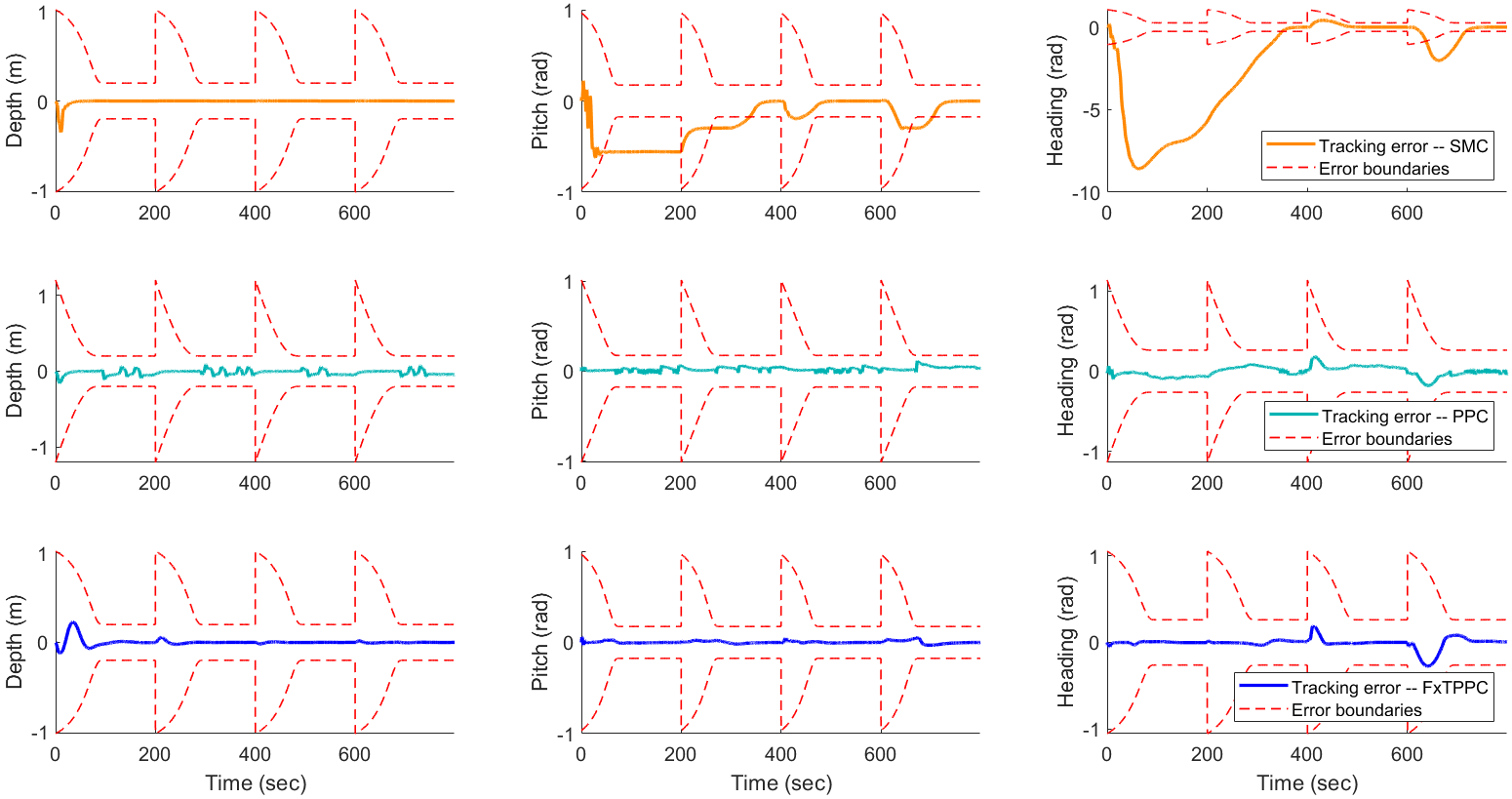}
    \caption{{Switching attitude tracking error comparison among SMC, PPC, and the proposed FxTPPC. SMC results in a large drift from the reference attitudes under the influence of disturbances. FxTPPC, compared with PPC, has smoother output tracking and more stable errors around zero. }}
    \label{fig: switch tracking error}
\end{figure*}

\begin{figure}
    \centering
    \includegraphics[width=.45\linewidth]{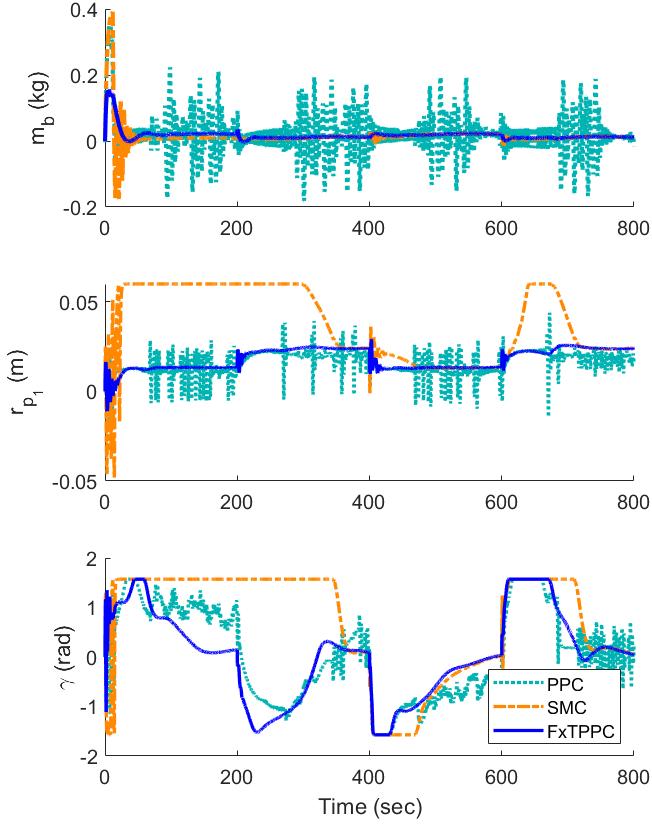}
    \caption{{Switching attitude tracking control effort comparison among SMC, PPC, and the proposed FxTPPC. The proposed controller results in smoother control efforts with lower chattering and shorter saturation endurance compared with the conventional methods. }}
    \label{fig: switch control effort}
\end{figure}

\begin{figure*}
    \centering
    \includegraphics[width=.8\textwidth]{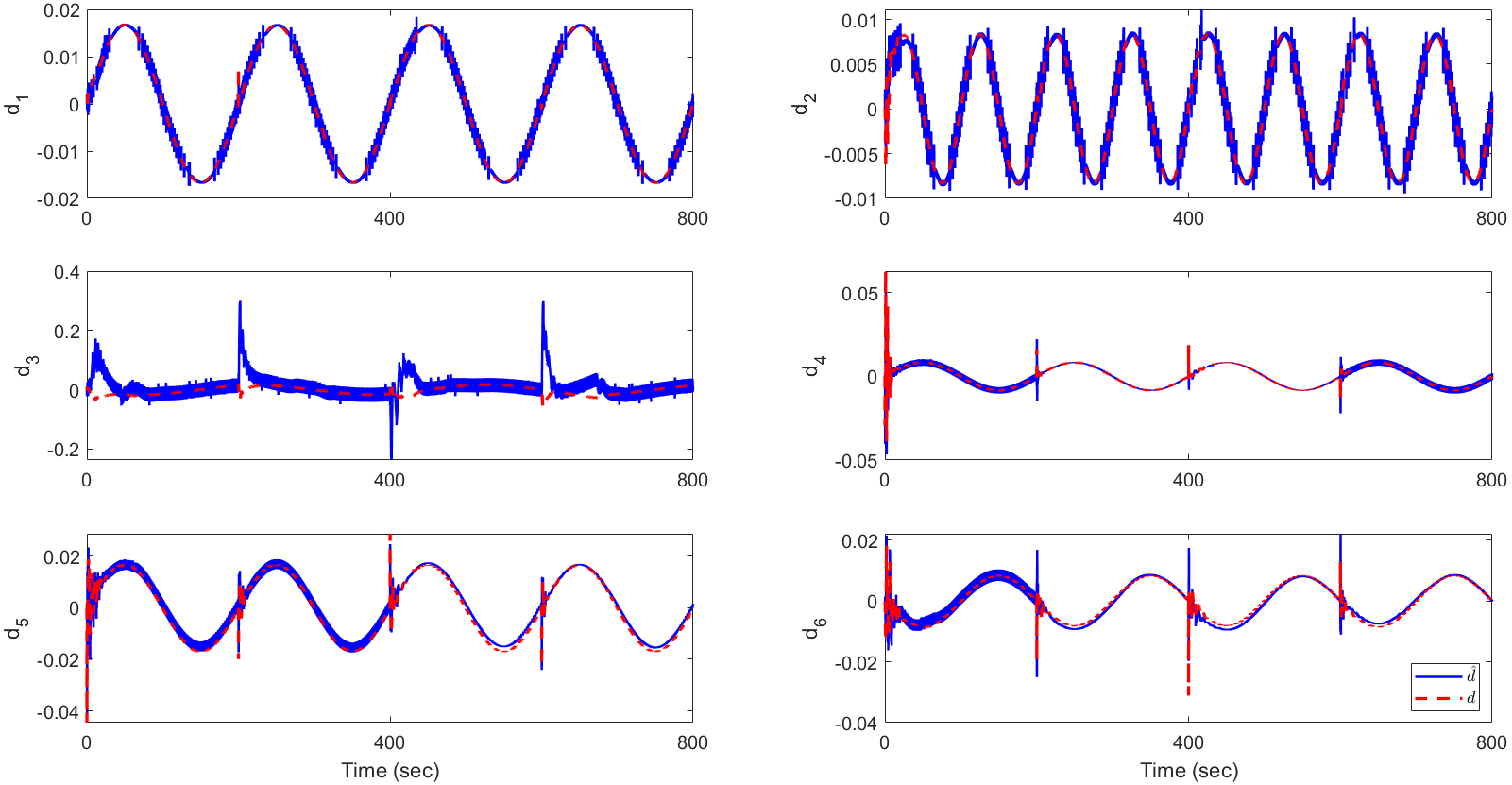}
    \caption{Estimate of the lumped disturbances in the body frame of the UG by the proposed disturbance observer. }
    \label{fig: switch disturbance estimate}
\end{figure*}

The switching attitude tracking performance is shown in Fig. \ref{fig: switch tracking performance}. The proposed method can track the periodically switched target heading angles under the influence of model uncertainties and environmental disturbances because of the proposed observer, while the SMC results in larger drifting from the desired attitudes. The tracking errors of all three control methods are compared in Fig. \ref{fig: switch tracking error}. The PPC results show large oscillations in the tracking errors and inconsistent steady state values. On the contrary, the proposed method leads to smooth and consistent tracking of the attitudes. Fig. \ref{fig: switch control effort} depicts that comparing to the SMC method, which has most of the control efforts at the physical limitation of the input actuators, and the PPC method, which results in large chattering issues in all three control inputs, the proposed FxTPPC method guarantees smooth control efforts within the input saturation bounds. Fig. \ref{fig: switch disturbance estimate} shows that the sliding mode disturbance observer can effectively estimate the time-varying lumped disturbances in the body frame of the UG when switching its attitudes. 

Table. \ref{tab: switch comparison} compares the control performance of the controllers for all control objectives. 
The control methods are evaluated by the following factors:
\begin{enumerate}
    \item Transient performance 
    \begin{equation}
        e_{i_M}=\max_t\{|e_i(t)|\}
    \end{equation}
    \item Average tracking performance
    \begin{equation}
        L_2[e_i]=\sqrt{\frac{1}{T_f}\int_0^{T_f}|e_i(t)|^2\:dt}
    \end{equation}
    \item Average control input
    \begin{equation}
        L_2[U_i] = \sqrt{\frac{1}{T_f}\int_0^{T_f}|u_i(t)|^2\:dt}
    \end{equation}
    \item Degree of control chattering
    \begin{equation}
        L_2[\Delta U_i] = \sqrt{\frac{1}{N}\sum_{j=1}^N|U_i(j\Delta T)-U_i((j-1)\Delta T)|^2}
    \end{equation}
\end{enumerate}
where $T_f$ and $N$ mark the final iteration of the simulation. The minimum of each comparison group is highlighted in the table. {As shown, the proposed fixed-time controller improves the average tracking performance by at least $95.41\%$ in pitch and $0.56\%$ in heading compared with the baseline control methods, SMC and PPC. It also decreases the control chattering by at minimum $96.01\%$ in depth, $92.11\%$ in pitch, and $56.76\%$ in heading.} Although FxTPPC is not superior in average depth and heading tracking compared with the other two controllers, it shows much lower average control efforts and chattering, which indicates that the proposed controller can consume less energy by minimizing the back-and-forth motions of the actuation hardware. 

Based on the analysis above, the proposed controller can achieve the preset performance in the presence of model uncertainties and environmental disturbances. The lumped disturbances can be estimated accurately in finite time by the proposed disturbance observer. Moreover, compared to the PPC method, the proposed controller can efficiently reduce the influence of control chattering.

\begin{table*}
\caption{Control performance comparison for target attitude following test case}
\centering
\ra{1.3}
\begin{adjustbox}{max width=\linewidth}
\begin{tabular}{@{}rrrrcrrrcrrr@{}}
\toprule& \multicolumn{3}{c}{\textbf{Depth}} & \phantom{abc}& \multicolumn{3}{c}{\textbf{Pitch angle}} &\phantom{abc} & \multicolumn{3}{c}{\textbf{Heading angle}}\\\cmidrule{2-4} \cmidrule{6-8} \cmidrule{10-12}& SMC & PPC & FxTPPC && SMC & PPC & FxTPPC && SMC & PPC & FxTPPC\\ 
\textbf{Transient} & 0.3502&0.2587&\cellcolor{lightgray}0.2273 && 0.6239&0.1153&\cellcolor{lightgray}0.0579 && 8.5853&\cellcolor{lightgray}0.1818&0.2732\\
\textbf{Average tracking} & \cellcolor{lightgray}0.0995&0.1130&0.1203 && 0.9783&0.1060&\cellcolor{lightgray}0.0449 && 11.6699&0.1977&\cellcolor{lightgray}0.1966\\
\textbf{Average control effort} &0.1382	&0.2319	&\cellcolor{lightgray}0.0793	&&0.1371	&0.0622	&\cellcolor{lightgray}0.0597	&&4.0989	&3.0909	&\cellcolor{lightgray}2.7979\\ 
\textbf{Degree of control chattering} &3.4095e-04	&0.0124	&\cellcolor{lightgray}1.3595e-05	&&5.0657e-05	&0.0017	&\cellcolor{lightgray}3.9991e-06	&&0.0592	&3.3109	&\cellcolor{lightgray}0.0256\\ 
\bottomrule
\end{tabular}
\end{adjustbox}
\label{tab: switch comparison}
\end{table*}

\subsection{Waypoint-based path following}\label{case2}
In this test case, the FxTPPC law is combined with the iLoS guidance law \eqref{eq: guidance law} for the glider to follow a series of waypoints. Each target waypoint is updated to the next once the following condition is met: 
\begin{equation}
    \sqrt{(X-X_{di})^2+(Y-Y_{di})^2}\leq R
\end{equation}
where $[X_{di},Y_{di}]^T$ is the current target waypoint in the horizontal plane, and $R\in\mathbb{R}>0$ is the waypoint switching distance. In this work, the guidance controller uses the following parameter values: $\Lambda=2.5$, $k_I=0.01$, $R=5$. 

The switching mechanism is added to the proposed FTPF \eqref{eq: PPF} when a new target waypoint is updated:
\begin{equation}
    \begin{split}
        t^*&=t-t_w \\ 
        P(t^*)& = \begin{cases}
        \mathrm{sech}(\mathrm{sech}(P_0)\cdot\frac{T}{T-t^*})+P_\infty, & 0\leq t^*<T \\ 
        P_\infty, &t^*\geq T
        \end{cases}
    \end{split}
\end{equation}
where $t_w$ is the time of new target updating. In this test case, the following parameter values are used for each FTPF: 
\begin{itemize}
    \item \textbf{$Z$} \text{--} $P_0=1$, $P_\infty=0.5$, $T=100$, $\delta_L=1$, $\delta_R=1$;
    \item \textbf{$\theta$} \text{--} $P_0=5\pi/18$, $P_\infty=\pi/18$, $T=80$, $\delta_L=1$, $\delta_R=1$;
    \item \textbf{$\psi$} \text{--} $P_0=5\pi/18$, $P_\infty=2\pi/45$, $T=60$, $\delta_L=1$, $\delta_R=1$. 
\end{itemize}
The performance function of PPC is set to have the same convergence setups as shown above. 

\begin{figure}
    \centering
    \includegraphics[width=.5\linewidth]{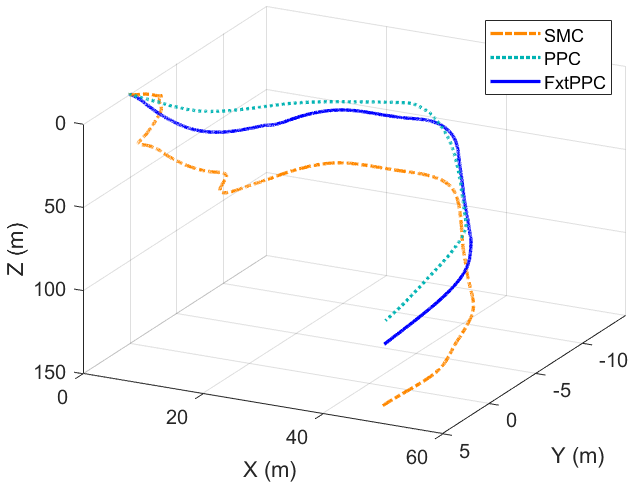}
    \caption{{Path following in 3D space, comparison among SMC, PPC, and the proposed FxTPPC. SMC results in a large drift, compared with the other two methods with disturbance observer, due to the impact of disturbances. }}
    \label{fig: waypoint 3d tracking}
\end{figure}

\begin{figure}
    \centering
    \includegraphics[width=.5\linewidth]{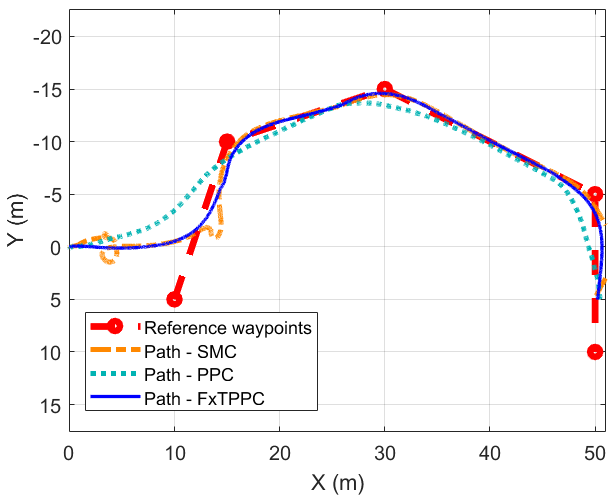}
    \caption{{Waypoint following in horizontal plane, comparison among SMC, PPC, and the proposed FxTPPC. The proposed method achieves the lowest drift from the preset course and more accurate tracking. }}
    \label{fig: waypoint 2d tracking}
\end{figure}

\begin{figure*}
    \centering
    \includegraphics[width=.8\textwidth]{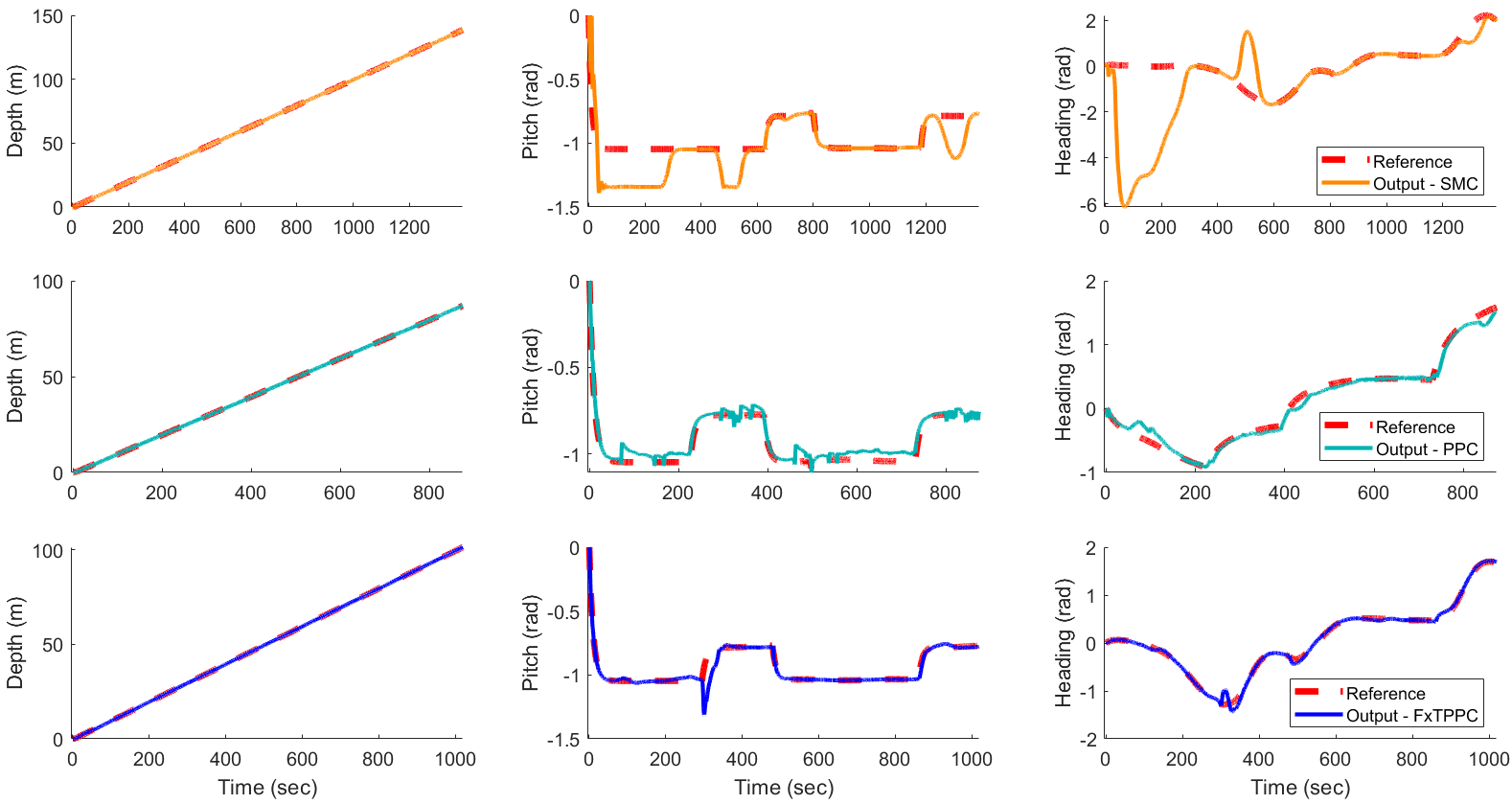}
    \caption{{Attitude tracking performance comparison among SMC, PPC, and the proposed FxTPPC. In the presence of lumped disturbances, SMC results in large drifting from the reference and PPC leads to high output chattering while FxTPPC is able to track the reference with high accuracy and low oscillation.}}
    \label{fig: waypoint tracking performance}
\end{figure*}

\begin{figure*}
    \centering
    \includegraphics[width=.8\textwidth]{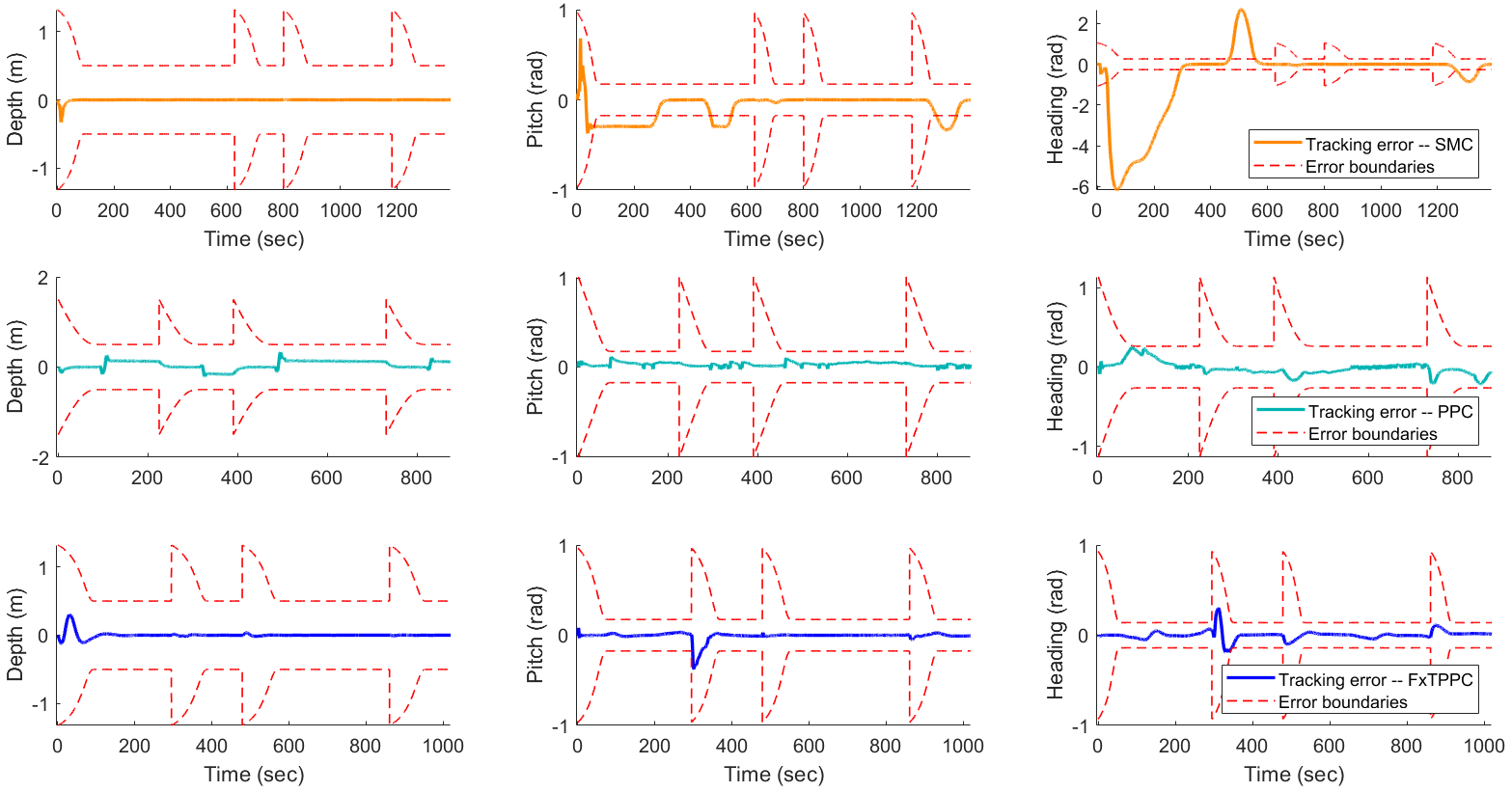}
    \caption{{Attitude tracking error comparison among SMC, PPC, and the proposed FxTPPC. SMC results in a large drift from the reference attitudes under the influence of disturbances. FxTPPC, compared with PPC, has smoother output tracking and more stable errors around zero.}}
    \label{fig: waypoint tracking error}
\end{figure*}

\begin{figure}
    \centering
    \includegraphics[width=.5\linewidth]{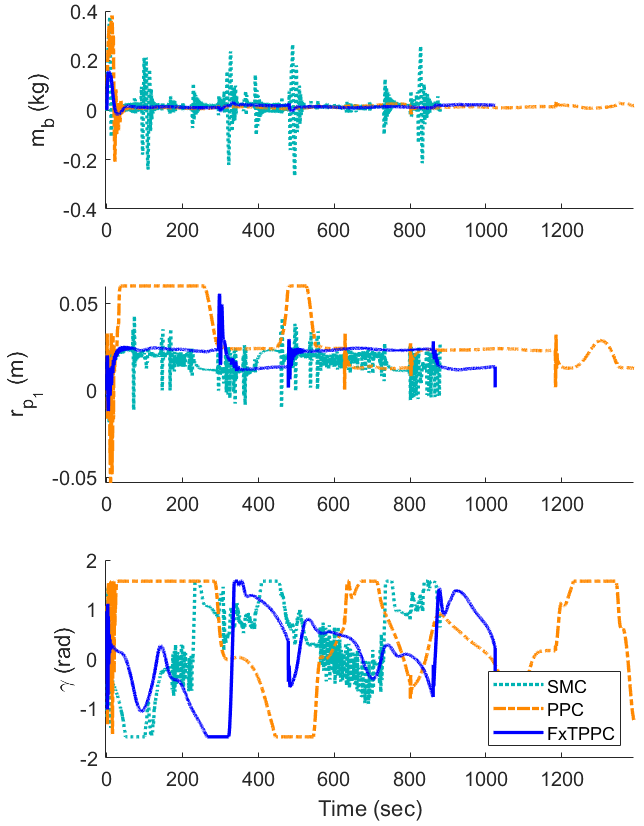}
    \caption{{Attitude tracking control effort comparison among SMC, PPC, and the proposed FxTPPC. The proposed controller results in smoother control efforts with lower chattering and shorter saturation endurance compared with the conventional methods. But PPC results in the fastest completion of the path tracking. }}
    \label{fig: waypoint control effort}
\end{figure}

\begin{figure*}
    \centering
    \includegraphics[width=.8\textwidth]{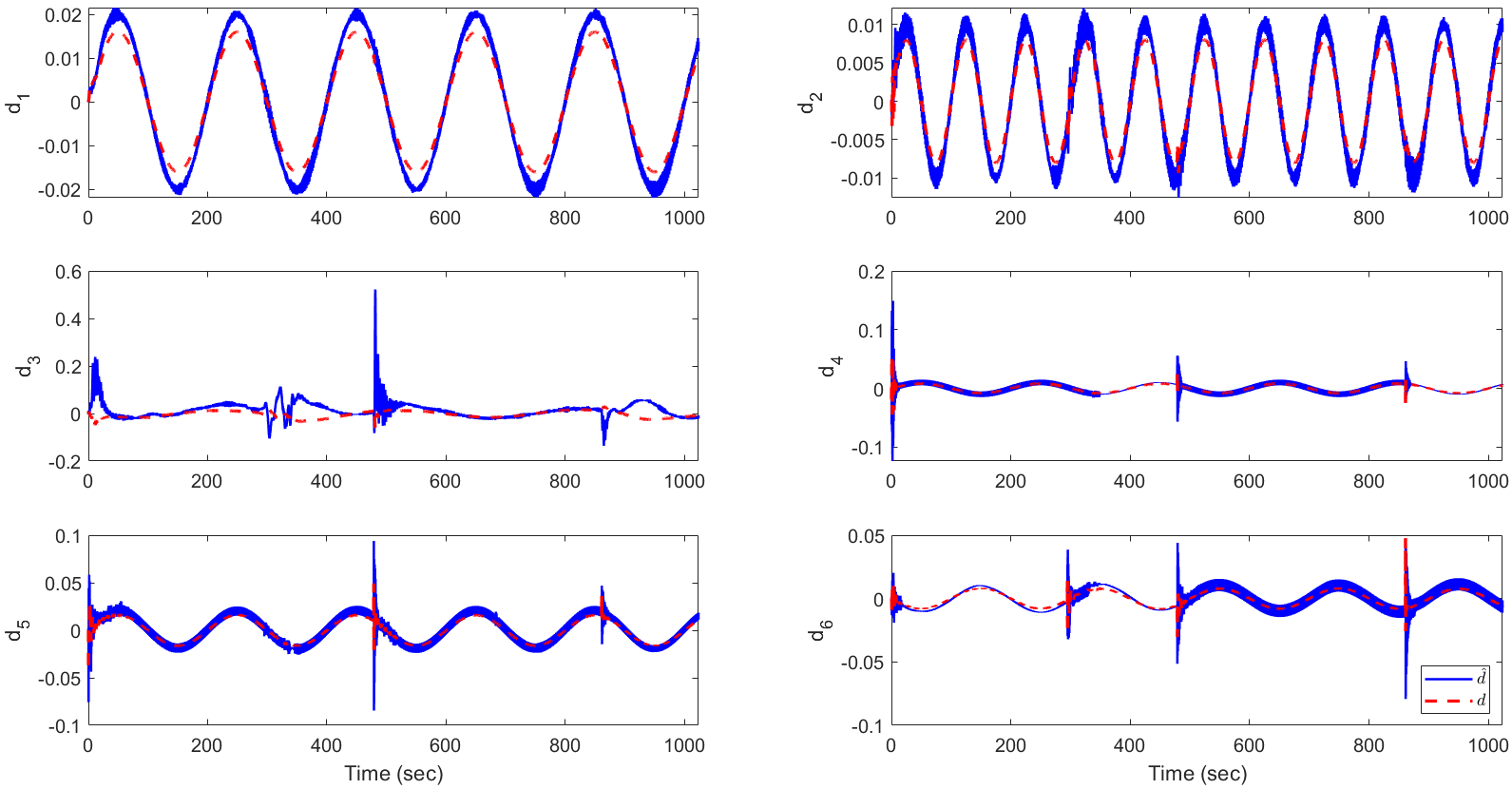}
    \caption{Estimate of the lumped disturbances in the body frame of the UG by the proposed disturbance observer. }
    \label{fig: waypoint disturbance estimate}
\end{figure*}

The path in the simulation is constructed with five waypoints: $[10,5]^T\rightarrow$ 
$[15,-10]^T\rightarrow$ 
$[30,-15]^T\rightarrow$ 
$[50,-5]^T\rightarrow$ 
$[50,10]^T$. The reference path in the vertical plane is defined by the gliding angles and depth in \eqref{eq: reference gliding angle} and \eqref{eq: reference depth}.  The vehicle initially starts at a position off the reference path. 
Fig. \ref{fig: waypoint 3d tracking} shows how the lack of disturbance estimation would lead the vehicle to drift from the desired position if controlled only by SMC. The waypoint tracking performance is illustrated in Fig. \ref{fig: waypoint 2d tracking}, which shows that FxTPPC can track the horizontal path while guaranteeing the smoothness and accuracy. SMC results in a large drift in the position, especially at the beginning of tracking. PPC can track the waypoints, but the position error along the path is more conspicuous compared with FxTPPC's result. Fig. \ref{fig: waypoint tracking performance} compares the attitude tracking performance among the three controllers. It is noticeable that FxTPPC leads to less drifting from the target attitude and lower chattering in the tracking compared to the other two controllers. This is further emphasized in Fig. \ref{fig: waypoint tracking error}, where the tracking errors are compared. SMC is unable to limit the tracking errors within the desired boundary. PPC controls the tracking errors to be stable within the prescribed performance, but has large chattering and inconsistent steady state values. In contrast to SMC and PPC, the proposed FxTPPC method results in smooth and consistent tracking errors while maintaining the prescribed performance. The control efforts of the three controllers are shown in Fig. \ref{fig: waypoint control effort}, in which FxTPPC has smoother control inputs with lower chattering, which leads to a safer control with lower energy consumption. It is also worth noting that the PPC takes the shortest time to finish the whole waypoint-tracking path, yet its chattering and drifting issues undermine its overall performance. Fig. \ref{fig: waypoint disturbance estimate} shows that the proposed observer can accurately track the lumped disturbances in finite time along the whole path following process. 

The controller performance, including the transient performance, average tracking performance, average control effort, and degree of control chattering, is evaluated in Table. \ref{tab: waypoint comparison}. {The proposed fixed-time controller improves the average tracking performance by at least $23.56\%$ in depth, $45.12\%$ in pitch, and $68.42\%$ in heading compared with the baseline control methods. It also decreases the control chattering by at minimum $54.58\%$ in depth, $74.27\%$ in pitch, and $43.11\%$ in heading.} Similar to the first test case, FxTPPC shows superiority in all three control objectives over SMC and PPC, even though PPC has better performance in the maximum transient of pitch and heading angles. 

Hence, the simulation results in this test case demonstrate that the proposed methodology in this paper effectively addresses the challenge of tracking straight-line paths and transiting waypoints in the presence of time-varying environmental disturbances and model uncertainties. This method is more robust in maintaining the vehicle's course and turning smoothly compared with the traditional SMC and PPC. 

\begin{remark}
{The structure of the proposed control scheme, comprising the fixed-time controller, the FTPF-based error transformation, and the fixed-time disturbance observer, relies primarily on algebraic computations and low-order system dynamics. This simplicity makes the method well-suited for deployment on the embedded platforms typically used in underwater gliders, which are often low-power single-board computers with limited computational resources. Real-time implementation can be achieved at modest update rates (e.g. $10-20 \si{Hz}$), fully compatible with the control cycles commonly employed in glider operations.}
\end{remark}


\begin{table*}
\caption{Control performance comparison for waypoint following test case}
\centering
\ra{1.3}
\begin{adjustbox}{max width=\linewidth}
\begin{tabular}{@{}rrrrcrrrcrrr@{}}
\toprule& \multicolumn{3}{c}{\textbf{Depth}} & \phantom{abc}& \multicolumn{3}{c}{\textbf{Pitch angle}} &\phantom{abc} & \multicolumn{3}{c}{\textbf{Heading angle}}\\\cmidrule{2-4} \cmidrule{6-8} \cmidrule{10-12}& SMC & PPC & FxTPPC && SMC & PPC & FxTPPC && SMC & PPC & FxTPPC\\ 
\textbf{Transient} & 0.3266	&0.3319	&\cellcolor{lightgray}0.2990	&&0.6817	&\cellcolor{lightgray}0.1143	&0.3776	&&6.1737	&\cellcolor{lightgray}0.2500	&0.3006\\
\textbf{Average tracking} & 0.1753	&0.3357	&\cellcolor{lightgray}0.1340	&&0.5086	&0.2200	&\cellcolor{lightgray}0.1516	&&5.9247	&0.2651	&\cellcolor{lightgray}0.1574\\
\textbf{Average control effort} &0.1152	&0.2126	&\cellcolor{lightgray}0.0689	&&0.1074	&0.0837	&\cellcolor{lightgray}0.0655	&&3.5468	&3.2404	&\cellcolor{lightgray}2.6476\\ 
\textbf{Degree of control chattering} &2.0991e-04	&0.0142	&\cellcolor{lightgray}1.1378e-05	&&3.1798e-05	&0.0017	&\cellcolor{lightgray}8.1838e-06	&&0.0566	&3.3337	&\cellcolor{lightgray}0.0322\\ 
\bottomrule
\end{tabular}
\end{adjustbox}
\label{tab: waypoint comparison}
\end{table*}

%%%%%%%%%%%%%%%%%%%%%%%%%%%%%%%%%%%%%%%%%%%%%%%%%%%%%%%%%%
\section{Conclusion}\label{Conclusion}
This paper presents a robust control scheme, fixed-time prescribed performance control, for 3D path following of underwater gliders. It adopts the finite-time performance function in combination with the fixed-time control law so that tracking error convergence within pre-defined transient and steady-state bounds may be ensured fast and uniformly, disregarding the initial conditions. {Simultaneously with fixed-time disturbance observer-based estimation of uncertainties and iLOS guidance law for waypoint tracking, it explicitly outperforms existing controllers in simulations regarding accuracy of tracking by at least $23.56\%$ in depth, $45.12\%$ in pitch, and $68.42\%$ in heading; it also decreases the control chattering by at minimum $54.58\%$ in depth, $74.27\%$ in pitch, and $43.11\%$ in heading.} FxTPPC therefore provides an energy-efficient solution for safe and robust UG navigation in the ocean environment.

For future work, a robust kinematic controller building upon the LoS method will be combined with the proposed FxTPPC to enable 3D waypoint tracking. This could provide UGs with a more robust and comprehensive solution for accurate 3D path following. Certain application scenarios may include moving target tracking and soft landing of UGs. In addition, though in this work it is assumed that all the system states are available as feedback to the control system, some UG models are not equipped with the sensors to provide these data in real-time. Thus, future work will also incorporate an extended state observer to handle sensor noise and unavailable velocity feedback, and explore formation control of multiple underwater gliders.

% \appendix
% \section{My Appendix}
% Appendix sections are coded under \verb+\appendix+.

% \verb+\printcredits+ command is used after appendix sections to list 
% author credit taxonomy contribution roles tagged using \verb+\credit+ 
% in frontmatter.
%%%%%%%%%%%%%%%%%%%%%%%%%%%%%%%%%%%%%%%%%%%%%%%%%%%%%%%%%%
\printcredits

\bibliographystyle{model1-num-names}
% \bibliographystyle{cas-model2-names}

\bibliography{cas-refs}

@misc{UGexample,
author={NOAA},
year={2024},
title={Physical Oceanography Division - Global Ocean Observations - NOAA/AOML - CARICOOS hurricane underwater gliders - previous missions}, 
url={https://www.aoml.noaa.gov/phod/goos/gliders/previous_missions.php#:~:text=Missions%2059%20through%2064:%20July%2DNovember%2C%202024%20(Hurricane,recovered%20by%20the%20end%20of%20November%202024.}, 
journal={Physical Oceanography Division - Global Ocean Observations - NOAA/AOML - CARICOOS hurricane underwater gliders - previous missions}
}

@ARTICLE{current_impact,
  
AUTHOR={von Oppeln-Bronikowski, Nicolai  and de Young, Brad  and Belzile, Melany  and Comeau, Adam  and Cyr, Frédéric  and Davis, Richard  and Emery, Pamela  and Richards, Clark  and Hebert, David  and Van Der Meer, Jude },
         
TITLE={Best practices for operating underwater gliders in Atlantic Canada},
        
JOURNAL={Frontiers in Marine Science},
        
VOLUME={Volume 10 - 2023},

YEAR={2023},

URL={https://www.frontiersin.org/journals/marine-science/articles/10.3389/fmars.2023.1108326},

DOI={10.3389/fmars.2023.1108326},

ISSN={2296-7745}}

@article{FTPFearly,
  title={Adaptive neural networks finite-time tracking control for non-strict feedback systems via prescribed performance},
  author={Liu, Yang and Liu, Xiaoping and Jing, Yuanwei},
  journal={Information Sciences},
  volume={468},
  pages={29--46},
  year={2018},
  publisher={Elsevier}
}

@inproceedings{PPC1st,
  title={Prescribed performance adaptive control of SISO feedback linearizable systems with disturbances},
  author={Bechlioulis, Charalampos P and Rovithakis, George A},
  booktitle={2008 16th mediterranean conference on control and automation},
  pages={1035--1040},
  year={2008},
  organization={IEEE}
}

@INPROCEEDINGS{yang2024underice,
  author={Yang, Hanzhi and Mahmoudian, Nina},
  booktitle={2024 IEEE/OES Autonomous Underwater Vehicles Symposium (AUV)}, 
  title={Under-ice Trajectory Tracking for Underwater Gliders Using Fuzzy-based Adaptive Control}, 
  year={2024},
  volume={},
  number={},
  pages={1-6},
  doi={10.1109/AUV61864.2024.11030787}}

@article{zhang2013spiraling,
  title={Spiraling motion of underwater gliders: Modeling, analysis, and experimental results},
  author={Zhang, Shaowei and Yu, Jiancheng and Zhang, Aiqun and Zhang, Fumin},
  journal={Ocean Engineering},
  volume={60},
  pages={1--13},
  year={2013},
  publisher={Elsevier}
}

@article{fossen1994guidance,
  title={Guidance and control of ocean vehicles},
  author={Fossen, Thor I},
  journal={(No Title)},
  year={1994}
}

@article{Tian2017,
   author = {Bailing Tian and Zongyu Zuo and Xiaomo Yan and Hong Wang},
   doi = {10.1016/J.AUTOMATICA.2017.01.007},
   issn = {0005-1098},
   journal = {Automatica},
   month = {6},
   pages = {17-24},
   publisher = {Pergamon},
   title = {A fixed-time output feedback control scheme for double integrator systems},
   volume = {80},
   url = {https://www.sciencedirect.com/science/article/pii/S000510981730016X},
   year = {2017}
}

@article{Polyakov2012,
   author = {Andrey Polyakov},
   doi = {10.1109/TAC.2011.2179869},
   issn = {00189286},
   issue = {8},
   journal = {IEEE Transactions on Automatic Control},
   pages = {2106-2110},
   title = {Nonlinear feedback design for fixed-time stabilization of linear control systems},
   volume = {57},
   year = {2012}
}

@article{Levant2005,
   author = {Arie Levant},
   doi = {10.1016/J.AUTOMATICA.2004.11.029},
   issn = {0005-1098},
   issue = {5},
   journal = {Automatica},
   month = {5},
   pages = {823-830},
   publisher = {Pergamon},
   title = {Homogeneity approach to high-order sliding mode design},
   volume = {41},
   url = {https://www.sciencedirect.com/science/article/pii/S0005109805000142},
   year = {2005}
}

@article{zhang2019fixed,
  title={Fixed-time extended state observer-based trajectory tracking and point stabilization control for marine surface vessels with uncertainties and disturbances},
  author={Zhang, Jingqi and Yu, Shuanghe and Yan, Yan},
  journal={Ocean Engineering},
  volume={186},
  pages={106109},
  year={2019},
  publisher={Elsevier}
}

@article{wang2023predictor,
  title={Predictor-based fixed-time LOS path following control of underactuated USV with unknown disturbances},
  author={Wang, Shandan and Sun, Mengwei and Xu, Yihang and Liu, Jian and Sun, Changyin},
  journal={IEEE Transactions on Intelligent Vehicles},
  volume={8},
  number={3},
  pages={2088--2096},
  year={2023},
  publisher={IEEE}
}

@article{su2021event,
  title={Event-triggered integral sliding mode fixed time control for trajectory tracking of autonomous underwater vehicle},
  author={Su, Bo and Wang, Hongbin and Li, Ning},
  journal={Transactions of the Institute of Measurement and Control},
  volume={43},
  number={15},
  pages={3483--3496},
  year={2021},
  publisher={SAGE Publications Sage UK: London, England}
}

@article{gao2020fixed,
  title={Fixed-time sliding mode formation control of AUVs based on a disturbance observer},
  author={Gao, Zhenyu and Guo, Ge},
  journal={IEEE/CAA Journal of Automatica Sinica},
  volume={7},
  number={2},
  pages={539--545},
  year={2020},
  publisher={IEEE}
}

@article{PPCexample4,
  title={Adaptive fuzzy controller design for dynamic positioning ship integrating prescribed performance},
  author={Wang, Yuanhui and Wang, Haibin and Li, Mingyang and Wang, Duansong and Fu, Mingyu},
  journal={Ocean Engineering},
  volume={219},
  pages={107956},
  year={2021},
  publisher={Elsevier}
}

@article{PPCexample7,
  title={Adaptive region tracking control with prescribed transient performance for autonomous underwater vehicle with thruster fault},
  author={Liu, Xing and Zhang, Mingjun and Wang, Suming},
  journal={Ocean Engineering},
  volume={196},
  pages={106804},
  year={2020},
  publisher={Elsevier}
}

@article{PPCexample6,
  title={Robust adaptive trajectory tracking control of underactuated autonomous underwater vehicles with prescribed performance},
  author={Li, Jian and Du, Jialu and Sun, Yuqing and Lewis, Frank L},
  journal={International Journal of Robust and Nonlinear Control},
  volume={29},
  number={14},
  pages={4629--4643},
  year={2019},
  publisher={Wiley Online Library}
}

@article{leonard2002model,
  title={Model-based feedback control of autonomous underwater gliders},
  author={Leonard, Naomi Ehrich and Graver, Joshua G},
  journal={IEEE Journal of oceanic engineering},
  volume={26},
  number={4},
  pages={633--645},
  year={2002},
  publisher={IEEE}
}

@article{UGPPC,
title = {Adaptive neural network sliding mode tracking control with prescribed performance for an underwater glider under input saturation},
journal = {Ocean Engineering},
volume = {307},
pages = {118150},
year = {2024},
issn = {0029-8018},
doi = {https://doi.org/10.1016/j.oceaneng.2024.118150},
author = {Xu Zhang and Baoheng Yao and Lian Lian and Zhihua Mao},
}

@article{Luo2025UGPPC,
   author = {Man Luo and Tianshu Wang and Rongshun Juan and Shoufu Liu and Junhe Wan and Zhongke Gao},
   doi = {10.1016/J.OCEANENG.2025.122304},
   issn = {0029-8018},
   journal = {Ocean Engineering},
   month = {11},
   pages = {122304},
   publisher = {Pergamon},
   title = {Fixed-time backstepping control of attitude tracking in 3D space for an autonomous underwater glider},
   volume = {340},
   url = {https://www.sciencedirect.com/science/article/pii/S0029801825019882#bib0052},
   year = {2025}
}

@article{yang2025UGPPC,
title = {Finite-time prescribed performance with fixed-time disturbance rejection for underwater glider heading control},
journal = {Ocean Engineering},
volume = {337},
pages = {121842},
year = {2025},
issn = {0029-8018},
doi = {https://doi.org/10.1016/j.oceaneng.2025.121842},
url = {https://www.sciencedirect.com/science/article/pii/S0029801825015483},
author = {Hanzhi Yang and Nina Mahmoudian},
}

@INPROCEEDINGS{iLoS,
  author={Borhaug, Even and Pavlov, A. and Pettersen, Kristin Y.},
  booktitle={2008 47th IEEE Conference on Decision and Control}, 
  title={Integral LOS control for path following of underactuated marine surface vessels in the presence of constant ocean currents}, 
  year={2008},
  volume={},
  number={},
  pages={4984-4991},
  doi={10.1109/CDC.2008.4739352}}

@inproceedings{PID2,
  title={Underwater glider motion control},
  author={Mahmoudian, Nina and Woolsey, Craig},
  booktitle={2008 47th IEEE Conference on Decision and Control},
  pages={552--557},
  year={2008},
  organization={IEEE}
}

@article{Cao2016Adaptive,
   author = {Junjun Cao and Junliang Cao and Zheng Zeng and Lian Lian},
   doi = {10.1177/1729881416669484/ASSET/9B5A7374-03C0-4EFD-951D-A916A9D0E09B/ASSETS/IMAGES/LARGE/10.1177_1729881416669484-FIG10.JPG},
   issn = {17298814},
   issue = {6},
   journal = {International Journal of Advanced Robotic Systems},
   month = {12},
   pages = {1-14},
   publisher = {SAGE Publications Inc.},
   title = {Nonlinear multiple-input-multiple-output adaptive backstepping control of underwater glider systems},
   volume = {13},
   url = {https://scholar.google.com/scholar_url?url=https://journals.sagepub.com/doi/pdf/10.1177/1729881416669484&hl=en&sa=T&oi=ucasa&ct=ufr&ei=5AaVaLi6H8jWieoPk8GJqAg&scisig=AAZF9b_9FxI6O8Neh8YYx6HyvKjO},
   year = {2016}
}

@article{Song2017ADRC,
   author = {Dalei Song and Tingting Guo and Hongdu Wang and Zhijian Cui and Liqin Zhou},
   doi = {10.1007/978-3-319-65289-4_69/TABLES/2},
   isbn = {9783319652887},
   issn = {16113349},
   journal = {Lecture Notes in Computer Science (including subseries Lecture Notes in Artificial Intelligence and Lecture Notes in Bioinformatics)},
   pages = {745-756},
   publisher = {Springer Verlag},
   title = {Pitch angle active disturbance rejection control with model compensation for underwater glider},
   volume = {10462 LNAI},
   url = {https://link.springer.com/chapter/10.1007/978-3-319-65289-4_69},
   year = {2017}
}

@article{Wang2022Review,
   author = {Jian Wang and Zhengxing Wu and Huijie Dong and Min Tan and Junzhi Yu},
   doi = {10.1109/JAS.2022.105671},
   issn = {23299274},
   issue = {9},
   journal = {IEEE/CAA Journal of Automatica Sinica},
   month = {9},
   pages = {1543-1560},
   publisher = {Institute of Electrical and Electronics Engineers Inc.},
   title = {Development and Control of Underwater Gliding Robots: A Review},
   volume = {9},
   year = {2022}
}

@inproceedings{abraham2015model,
  title={Model predictive control of buoyancy propelled autonomous underwater glider},
  author={Abraham, Ian and Yi, Jingang},
  booktitle={2015 American Control Conference (ACC)},
  pages={1181--1186},
  year={2015},
  organization={IEEE}
}

@inproceedings{zhang2014three,
  title={Three-dimensional spiral tracking control for gliding robotic fish},
  author={Zhang, Feitian and Tan, Xiaobo},
  booktitle={53rd IEEE Conference on Decision and Control},
  pages={5340--5345},
  year={2014},
  organization={IEEE}
}


\end{document}